\documentclass{aa}
\usepackage{graphicx} 
\usepackage{txfonts}
\begin{document}
\title{Did most present-day spirals form during the last 8 Gyrs?}

   \subtitle{A formation history with violent episodes revealed by panchromatic observations}

\author{  F. Hammer\inst{1}
          \and
          H. Flores\inst{1}
          \and
          D. Elbaz\inst{2}
          \and
	  X.Z. Zheng\inst{1,3}
          \and
	  Y.C. Liang\inst{1,4}
          \and
          C. Cesarsky\inst{5}
          }

\offprints{francois.hammer@obspm.fr}
\authorrunning{F. Hammer et al.}
\titlerunning{A recent formation of most spirals? }

\institute{ Laboratoire Galaxies, Etoiles, Physique et Instrumentation,
          Observatoire de Paris, 92195 Meudon, France
         \and
       CEA, Saclay-SAp, Ormes des Merisiers, 91191 Gif surYvette, France
       \and
       Max-Planck Institut f$\ddot{u}$r Astronomie, K$\ddot{o}$nigstuhl 17, D-69117 Heidel
berg, Germany
       \and
       National Astronomical Observatories, CAS, Beijing 100012, China
       \and
       ESO, Karl-Schwarzschild Strasse 2, D85748 Garching bei Muenchen, Germany
        }
\date{Received 15 June 2004 / Accepted 20 October 2004}

\abstract{
Studies of distant galaxies have shown that ellipticals and large spirals (Schade et al. 1999; Lilly et al. 1998) were already in place 8 Gyr ago, leading to a very modest recent star formation (Brinchmann and Ellis, 2000) in intermediate mass galaxies (3-30 $10^{10}$ $M_{\odot}$). This is challenged by a recent analysis (Heavens et al. 2004) of the fossil record of the stellar populations of $\sim$ $10^{5}$ nearby galaxies, which shows that intermediate mass galaxies formed or assembled the bulk of their stars 4 to 8 Gyr ago. Here we present direct observational evidence supporting this findings from a long term, multi-wavelength study of 195 z$>$ 0.4 intermediate mass galaxies, mostly selected from the Canada France Redshift Survey (CFRS). We show that recent and efficient star formation is revealed at IR wavelengths since $\sim$ 15\% of intermediate mass galaxies at z$>$ 0.4 are indeed luminous IR galaxies (LIRGs), a phenomenon far more common than in the local Universe. The star formation in LIRGs is sufficient in itself to produce 38\% of the total stellar mass of intermediate mass galaxies and then to account for most of the reported stellar mass formation since z=1. \\ 
Observations of distant galaxies have also the potential to resolve their star formation and mass assembly histories. The high occurrence of LIRGs is easily understood only if they correspond to episodic peaks of star formation, during which galaxies are reddened through short IREs (infrared episodes).  We estimate that each galaxy should experience 4 to 5 $\times$ $(\tau_{\rm IRE}/0.1Gyr)^{-1}$ IREs from z=1 to z=0.4, $\tau_{\rm IRE}$ being the characteristic timescale. 
An efficient and episodic star formation is further supported by the luminosity-metallicity relation of z$\sim$ 0.7 emission line galaxies, which we find to be on average metal deficient by a factor of $\sim$ 2 when compared to those of local spirals. 
We then examine how galaxy IREs can be related to the emergence at high redshift of the abundant population of galaxies with small size (but not with small mass), blue core and many irregularities. We show that recent merging and gas infall naturally explain both morphological changes and episodic star formation history in a hierarchical galaxy formation frame.\\
We propose a simple scenario in which 75$\pm$25\% of intermediate mass spirals have recently experienced their last major merger event, leading to a drastic reshaping of their bulges and disks during the last 8 Gyrs. It links in a simple manner distant and local galaxies, and gives account of the simultaneous decreases during that period, of the cosmic star formation density, of the merger rate, and of the number densities of LIRGs, compact and irregular galaxies, while the densities of ellipticals and large spirals are essentially unaffected. It predicts that 42, 22 and 36\% of the IR (episodic) star formation density is related to major mergers, minor mergers and gas infall, respectively.   
     
       }

\maketitle
%

\section{Introduction}
Understanding how and when galaxies and the stars in them were formed is still a considerable challenge for astrophysicists. A global picture of galaxy evolution during the last 8 Gyrs has emerged from the results of all-inclusive galaxy surveys.  The apparently unchanged number density of ellipticals and of large spirals at z$\sim$ 0.7 (Schade et al. 1999; Lilly et al. 1998; Conselice et al, 2004a) has lead to a picture for which most of the recent star formation  occurred in small mass galaxies (Cowie et al. 1996; Brinchman and Ellis, 2000, hereafter BE2000). This picture has been called "galactic downsizing" to illustrate its apparent contradiction to expectations from hierarchical models of galaxy formation. It is only consistent with a modest fraction of stars recently formed, since present-day stars are mostly locked in spheroids (bulge of spirals and ellipticals, Fukugita, Hogan \& Peebles, 1998). 

 This is challenged by the evolution of the cosmic stellar mass density which is based on a detailed modelling of galaxy spectral energy distributions (Cole et al. 2001; Dickinson et al. 2003,; Fontana et al. 2003; Drory et al. 2004; Rudnick et al. 2003). Even if they suffer from various uncertainties (cosmic variance, assumptions on stellar populations, metal abundances, extinction and initial mass function), all these studies predict that 30 to 50\% of the mass locked in stars in present day galaxies actually condensed into stars at $z<1$. These estimates agree with integrations of the universal star formation history, if it properly accounts for the light re-emitted by dust at IR wavelengths (Flores et al. 1999; Chary and Elbaz, 2001). Heavens et al. (2004, hereafter, H2004) have derived a peak of star formation at modest redshifts (z=0.5-0.6), mostly due to intermediate mass galaxies (3-30 $10^{10}M_{\odot}$) which have assembled/formed 60\% of their stellar masses during the last 8 Gyrs (from integrations of their Figure 2). 
There is now a reasonable agreement that a significant part of the present-day stars (30\% to 50\%) was formed at recent epochs. 

Where are these relatively recent stars born and how they have been formed ? We examine the major evolutionary trends since z=1 from all inclusive surveys. Five to eight Gyrs ago, the galaxy population included $ 70$\% starbursts ( defined as W$_O$([OII])$> 15\AA$ ), a fraction which has sharply decreased to 17\% in the local Universe (e.g. Hammer et al. 1997). Both UV (Lilly et al. 1996) and IR (Flores et al. 1999) comoving luminosity densities have been declining rapidly since $z=1$.

\subsection{ Most of the star formation since z=1 is related to the IR luminosity density}  

At IR wavelengths the comoving luminosity density was 70 times larger at $z=1$ than it is today (Elbaz et al. 1999; Elbaz and Cesarsky, 2003). At $z=1$ it is mostly due to luminous IR galaxies (LIRGs, defined as L$_{IR} > 10^{11} L_\odot$ following Sanders and Mirabel, 1996), which are mainly powered by star formation  (Fadda et al 2002). LIRGs form stars at very high rates (from  20 to several hundred of $M_\odot/yr$) and are responsible for the bulk of the cosmic infrared background (Elbaz et al. 2002). An early conclusion from studies of mid infrared emission and SEDs of galaxies is that from 50 to 66\% of the CSFD at $z\sim 0.7$ is related to far infrared emission (Flores et al. 1999), mostly due to bright mid-IR sources (LIRGs). The large impact of star formation revealed by the IR has been recently confirmed by Spitzer counts of 24 $\mu$m sources (Papovich et al. 2004).

Distant LIRG morphologies encompass those of disks, irregulars and LCGs, with a noticeable fraction of them showing signs of merging or interactions (Zheng et al. 2004a). If sustained at the observed high rates, their star formation can double their mass in less than 1 Gyr (Franceschini et al. 2003; Zheng et al. 2004a). Liang et al. (2004b) have derived their extinction, star formation rate and metal abundances from a detailed analysis of their high quality spectra from VLT. Distant LIRGs present a metal content less than half of that of local bright disks. Liang et al. (2004b) and Zheng et al. (2004a) have suggested that many distant disks are still forming a significant fraction of their stellar mass content.
   
\subsection{ Can we relate the UV luminosity density decline to the evolution of the merging rate ?}

Peculiar galaxies are found to be the most rapidly evolving galaxy population and to be responsible for most of the decline of the UV luminosity density (Brinchman et al. 1998; Lilly et al. 1998, BE2000, Conselice et al, 2004b). Besides showing unusual or irregular morphologies, peculiar galaxies generally display small half light radii (Lilly et al. 1998), while their stellar masses are comparable to those of large, more regular galaxies (BE2000). It has been often suggested that the morphologies of peculiar galaxies are related to interactions and that merging could play a major role in their number density decline. This is broadly consistent with the large expected amount of merging events during the last 8 Gyrs. From number counts of pairs up to $z=1$, Le F\`evre et al. (2000) estimated that each luminous (L*) galaxy experiences 1.1 to 1.8 $\times$ (7 10$^8$ yrs/$\tau_{merger}$) major merging event (with a companion with half and one fourth of its luminosity, respectively). In this study the characteristic time over which the merger is completed, $\tau_{merger}$ was assumed to range from 0.4 to 1 Gyr following expectations from simulations (Hernquist \& Mihos, 1995). A recent study made in the near IR (Bundy et al. 2004) leads us to suspect a slower merger history, and accounting for the various selection effects (colour \& relative spatial resolution) would decrease the Le F\`evre et al. estimates to half their original values. Merging appears still to be an important ingredient in galaxy evolution at z$<$ 1, and this matches rather well with a recent stellar mass production.

Several clues also favor a merger origin for many peculiar galaxies. In the CFRS, Hammer et al. (2001) have studied in detail the most compact ones ($r_{50}<$ 3.5 kpc), hereafter called luminous compact galaxies (LCGs) which show blue U-V colours but not necessarily blue B-K colours. LCGs are found to be dusty enshrouded starbursts (SFR averaged to 40 M$_{\odot}/yr$) superimposed on a mix of metal-rich old stellar population and intermediate age stars. Hammer et al (2001) suggested that LCGs could be remnants of mergers or interactions, and could be an important stage in the bulge formation in many spirals. Detailed studies of few local counterparts by Barton \& Zee (2001) and by Ostlin et al. (2001) have revealed tidally triggered central starbursts or complex velocity fields typical of completed mergers, which also explain their narrow emission lines. There are also theoretical grounds which support that merger remnants could be observed in large numbers in the distant Universe. Although after a major merger, the dynamical relaxation time is short (few $10^{7}$ yrs, Barnes \& Hernquist, 1992), hydro-dynamical models (Tissera et al. 2002; see also Baugh et al. 1996) show that enhanced star formation can be sustained from several $10^{8}$ yrs to $\sim$ 2 $10^{9}$ yrs after the collision, a period during which formation of stars in the surrounding disk can be accommodated.

\subsection{The present study}
 
Thus there are good reasons to suspect that merging probably has an important impact on the observed recent formation of stars. Because a significant part of the recent star formation is related to dust-enshrouded galaxies, it calls for a panchromatic study of galaxy properties. This study is based on long-term follow-up observations of the CFRS by our team, using HST, ISO, VLA and VLT. It has the goal to link the morphological changes seen in optical wavelengths to the star formation revealed at IR wavelengths, and, more ambitiously, to be able to see an overall picture of galaxy evolution over the last 8 Gyr. It targets 195 $z> 0.4$ galaxies, mostly from the sample of the CFRS which is probably the most widely studied survey of galaxies for its statistical properties. CFRS is essentially complete up to z$\sim$ 1, encompassing all luminous ($M_B<-20$) galaxies (Hammer et al. 1997) with stellar masses from 3 to 30 10$^{10} M_{\odot}$, those being responsible for the bulk of the star formation history reported by H2004. There is a general consensus that intermediate mass galaxies dominate the stellar mass budget at $z < 1$ (80\% of the total mass according to BE2000, 65\% from integrations of H2004). 

In Section 2 we present the data and a comparison of the various estimates of star formation rate is given in Section 3. A solid estimate of the stellar mass fraction formed in the past 8 Gyrs is provided in Section 4, as well as new insights on the star formation histories of individual galaxies. Section 5 describes the major morphological changes during this elapsed time, which are linked to the galaxy star formation history in Section 6, leading to a simple scenario which accounts for all changes in the intermediate mass galaxy population, in the framework of the hierarchical galaxy formation theory. $\Lambda$CDM cosmology 
($H_0$=70 km$s^{-1}Mpc^{-3}$, $\Omega_M=0.3$ and $\Omega_\Lambda=0.7$) is adopted throughout the paper.

\section{The data obtained from our panchromatic study of distant galaxies}

 This  randomly selected sample of galaxies (for a detailed description, see Liang et al, 2004b; Zheng et al, 2004a) was contructed in order to study each subclass of galaxies (LIRGs, LCGs, disk galaxies and starbursts). To study LIRGs two additional ISOCAM fields were included to avoid any contamination due to cosmological variance: selection and imagery on extra deep ISOCAM fields were done in the same way than for the CFRS survey ($I_{AB}< 22.5$ galaxies).\newline 
Table 1 summarizes the properties of 195 $I_{AB}$ $\le$ 22.5 galaxies with $z> 0.4$, on which 163, 14 and 23 are collected from the CFRS, the UDSR and the UDSF fields, respectively. Redshift measurements (and derived absolute magnitudes in the AB system) have been derived from CFRS or from VLT spectroscopy. The VLT follow-up was based on optical and near IR spectroscopy\footnote{FORS1 (65.0-352(A) and 68A-0298((A) runs), FORS2 (66.A-0599(A,B), 67A-0218(A) and 68A-0382(A)
runs) and ISAAC (68A-0298((B) and 69B-0301(A) runs)} at resolution R$> 1000$, aiming at studying properties of $z > 0.4$ galaxies in the fields of the southern hemisphere (all fields but the CFRS14h). The new data gathered here allow us to compare the physical properties of the most evolving galaxies (LIRGs and LCGs) to those of the rest of the galaxy population detected by the CFRS. In the following, conclusions relative to the general galaxy population will be systematically derived from statistics based on the whole CFRS sample. 

The HST follow-up was based on numerous WFPC2 archival data on the CFRS 03h and CFRS 14h fields on which we select galaxies with deep images in two colours (mainly F606W and F814W, see Zheng et al. 2004a for details). Central colours and compactness are derived from multiple color HST images (see Zheng et al. 2004a).

Stellar masses of galaxies can be estimated from their near IR luminosities (Charlot, 1999; Bell and de Jong, 2000). Here we have chosen a conservative approach, which assumes that $M/L_K$ depends on the rest-frame B-V color following the relationship derived by Bell et al (2003). We verified that an alternative use of the Bell \& de Jong (2000) relationship would not much affect our results and typical uncertainties are 0.1-0.2 dex (Bell et al, 2003). Rest frame colors and absolute K band magnitudes were derived from observed V-I and I-K (and K magnitude), respectively. This leads to stellar mass values in the same range as those derived by BE2000 for the CFRS sample: most stellar masses of z$>$ 0.4 CFRS galaxies range from 10.5 to 11.5 on a logarithmic scale. 

Four fields (CFRS 03h, 14h, UDSR and UDSF) were observed by deep ISOCAM pointings (Flores et al. 1999; 2004; Elbaz et al 2004) providing measurements of 15$\mu$m luminosities which have been converted in IR luminosities (see Elbaz et al. 2002). Star formation rates based on 3 different luminosities (IR, $H\alpha$, [OII]$3727$) assume the same IMF (see Hammer et al. 2003 and section 3). $H\alpha$ luminosities are derived from VLT measurements (see Flores et al. 2004; Liang et al. 2004b) while some [OII]$3727$ luminosities are derived from CFRS spectra after a proper account for aperture corrections (see Liang et al. 2004b).  $O/H$ gas abundances are calculated (see Liang et al. 2004b) from the ratio of [OII]$3727$ and [OIII]$5007$ lines to $H\beta$ after a
proper account for extinction effects. A few values of $O/H$ are however directly derived from the single [OIII]$5007$/$H\beta$ ratio (see Liang et al. 2004b).

\begin{table*}
\caption { Sample of distant galaxies }
{\scriptsize
 \begin{tabular}{cccccrrrcccl} \hline
CFRS & z & M$_{B}$  & M$_{K}$ &log(M/M$_{\odot}$)& SFR$_{\rm IR}^a$ &  SFR$_{H\alpha}^a$ &SFR$_{[\rm OII]}^a$ & 12+log(O/H) & C.C.$^b$ & Comp$^c$ & Morph$^d$ \\ \hline
00.0137 &  0.9512 &  -21.91 &  -23.15 &  10.98  &     ---     &  74.28    &  93.48   &  8.40   &  ---  &  --- &   ---  \\
00.0141 &  0.4401 &  -20.84 &  -21.49 &  10.36  &     ---     &   3.58    &   ---    &  8.60   &  ---  &  --- &   ---  \\
00.0564 &  0.6105 &  -19.79 &  -21.61 &  10.44  &     ---     &   1.15    &   1.61   &   ---   &  ---  &  --- &   ---  \\
00.0763 &  0.8360 &  -19.50 &  -21.16 &  10.32  &     ---     &   2.92    &   2.85   &   ---   &  ---  &  --- &   ---  \\
00.0868 &  0.7378 &  -20.63 &    ---  &   ---   &     ---     &  16.62    &   ---    &  8.71   &  ---  &  --- &   ---  \\
00.1175 &  0.6390 &  -20.23 &    ---  &   ---   &     ---     &  11.81    &   ---    &  8.74   &  ---  &  --- &   ---  \\
00.1721 &  0.5581 &  -19.95 &  -21.09 &  10.20  &     ---     &  13.36    &   ---    &  8.58   &  ---  &  --- &   ---   \\
03.0006 &  0.8836 &  -18.36 &  -19.59 &   9.58  &     69.65   &   ---     &   1.69   &   ---   &  ---  &  --- &   ---   \\ \hline
03.0032 &  0.5220 &  -19.61 &  -22.26 &  10.76  &     N/D     &   ---     &   1.73   &   ---   & 2.154 & 0.62 & spiral     \\
03.0035 &  0.8804 &  -21.12 &  -23.11 &  11.07  &    178.30   &   ---     &   3.30   &   ---   & 1.685 & 0.39 & spiral    \\
03.0046 &  0.5120 &  -19.76 &  -21.67 &  10.50  &     N/D     &   ---     &   3.93   &   ---   & 1.713 & 0.25 & spiral     \\
03.0062 &  0.8252 &  -21.28 &  -22.68 &  10.87  &   125.40    &   ---     &   4.32   &   ---   & 1.543 & 0.36 & spiral   \\
03.0085 &  0.6100 &  -19.42 &  -20.80 &  10.10  &     50.88   &   ---     &   2.30   &   ---   & 1.466 & 0.31 & spiral    \\
03.0174 &  0.5250 &  -18.52 &  -20.87 &  10.16  &     27.98   &   ---     &   0.70   &   ---   &  ---  & ---  &  ---       \\
03.0186 &  0.5220 &  -19.11 &  -21.95 &  10.61  &     26.39   &   8.83    &   1.66   &   ---   &  ---  & ---  &  ---       \\
03.0327 &  0.6064 &  -19.80 &  -21.41 &  10.35  &     N/D     &   ---     &   3.55   &   ---   & 1.539 & 0.64 & compact   \\
03.0350 &  0.6950 &  -20.84 &  -22.51 &  10.84  &     N/D     &   ---     &   ---    &   ---   & 2.005 & 0.58 &  e/s0     \\
03.0422 &  0.7150 &  -20.91 &  -22.47 &  10.76  &     114.20   &   71.75   &   4.75   &   ---   &  ---  & ---  &  ---       \\
03.0442 &  0.4781 &  -19.56 &  -20.40 &   9.93  &     N/D     &   2.91    &   3.34   &  7.92   &  ---  & ---  &  ---     \\
03.0445 &  0.5300 &  -20.34 &    ---  &   ---   &     25.67   &   6.05    &   3.05   &  8.72   & 1.706 & 0.23 & spiral   \\
03.0480 &  0.6080 &  -19.33 &  -19.62 &   9.59  &     75.89   &   ---     &  12.60   &   ---   &  ---  &  --- &  ---     \\
03.0507 &  0.4660 &  -20.19 &  -21.04 &  10.17  &     19.92   &   35.49   &   3.56   &  8.55   & 1.706 & 0.46 & spiral   \\
03.0508 &  0.4642 &  -19.55 &  -20.31 &   9.86  &     N/D     &   ---     &   6.75   &   ---   & 1.269 & 0.45 & compact   \\
03.0523 &  0.6508 &  -20.53 &  -21.52 &  10.36  &     99.66   &   79.73   &  11.33   &  8.61   & 1.070 & 0.54 & compact  \\
03.0528 &  0.7140 &  -20.68 &  -22.48 &  10.79  &     N/D     &   ---     &   ---    &   ---   & 1.915 & 0.43 & spiral    \\
03.0533 &  0.8290 &  -21.10 &  -22.41 &  10.73  &     189.85  &   ---     &   9.79   &   ---   & 1.339 & 0.38 & irregular\\
03.0560 &  0.6968 &  -20.66 &    ---  &   ---   &     N/D     &   ---     &   3.23   &   ---   & 2.086 & 0.42 & e/s0      \\
03.0570 &  0.6480 &  -19.94 &  -20.39 &   9.92  &     51.60   &   11.62   &   4.50   &  8.79   & 1.180 & 0.46 & compact  \\
03.0589 &  0.7160 &  -20.06 &  -21.08 &  10.21  &     N/D     &   ---     &   3.88   &   ---   & 1.200 & 0.43 & compact   \\
03.0595 &  0.6044 &  -20.15 &  -21.44 &  10.35  &     N/D     &   23.60   &   3.30   &   ---   &  ---  & ---  & ---      \\
03.0603 &  1.0480 &  -22.68 &  -24.40 &  11.48  &     N/D     &   ---     &   ---    &   ---   & 0.443 & 0.61 & compact   \\
03.0615 &  1.0480 &  -21.45 &  -22.03 &  10.56  &     N/D     &   ---     &  19.63   &   ---   & 1.168 & 0.47 & compact   \\
03.0619 &  0.4854 &  -20.62 &  -21.93 &  10.50  &     N/D     &   ---     &   1.56   &   ---   & 1.718 & 0.45 & spiral    \\
03.0645 &  0.5275 &  -20.20 &  -21.34 &  10.28  &     N/D     &   14.85   &   8.98   &  8.64   & 1.094 & 0.51 & irregular \\
03.0717 &  0.6070 &  -20.56 &    ---  &   ---   &     N/D     &   ---     &   1.23   &   ---   & 1.947 & 0.30 & spiral    \\
03.0728 &  0.5200 &  -18.96 &    ---  &   ---   &     N/D     &   ---     &   ---    &   ---   & 1.976 & 0.63 & e/s0      \\
03.0767 &  0.6690 &  -19.89 &  -22.45 &  10.81  &     N/D     &   ---     &   2.34   &   ---   & 2.132 & 0.63 & e/s0      \\
03.0776 &  0.8830 &  -20.35 &  -20.65 &  10.02  &     111.7   &   96.17   &   6.07   &   ---   & 1.374 & 0.56 & compact   \\
03.0837 &  0.8215 &  -20.44 &  -23.49 &  11.23  &     N/D     &   ---     &   8.51   &   ---   & 2.264 & 0.59 & irregular \\ 
03.0844 &  0.8220 &  -20.68 &  -22.87 &  10.97  &     N/D     &   ---     &   ---    &   ---   & 2.078 & 0.67 & e/s0       \\      
03.0916 &  1.0300 &  -21.77 &    ---  &   ---   &     N/D     &   ---     &   ---    &   ---   &  ---  &  --- & compact   \\
03.0932 &  0.6478 &  -20.12 &  -22.82 &  10.92  &     167.3   &   194.1   &   4.41   & 8.73    & 1.865 & 0.20 & spiral    \\
03.1032 &  0.6180 &  -21.14 &  -22.61 &  10.86  &     111.3   &   ---     &   4.01   &   ---   &  ---  & ---  &  ---        \\
03.1175 &  0.5610 &  -19.49 &  -20.84 &  10.15  &     68.20   &   ---     &   ---    &   ---   &  ---  & ---  &  ---        \\
03.1242 &  0.7690 &  -20.50 &  -22.17 &  10.62  &     86.55   &    82.20  &   9.72   &   ---   &  ---  & ---  &  ---      \\
03.1309 &  0.6170 &  -20.89 &  -22.91 &  10.92  &     153.7   &    56.08  &   5.02   & 8.37    &  ---  &  --- & merger    \\
03.1318 &  0.8575 &  -20.60 &  -21.01 &  10.14  &     N/D     &   ---     &  14.16   &   ---   &  ---  &  --- & merger     \\
03.1345 &  0.6167 &  -20.40 &  -22.95 &  10.98  &     N/D     &    16.7   &  12.70   &   ---   & 1.857 & 0.28 & spiral     \\
03.1349 &  0.6155 &  -20.97 &  -22.92 &  10.94  &     97.81   &    40.35  &   2.60   & 8.69    & 1.388 & 0.39 & spiral    \\
03.1353 &  0.6340 &  -20.23 &  -22.43 &  10.76  &     N/D     &   ---     &   4.07   &   ---   & 1.809 & 0.35 & spiral     \\
03.1522 &  0.5870 &  -19.49 &  -22.19 &  10.71  &     79.42   &   ---     &   ---    &   ---   & 2.260 & 0.36 & spiral    \\
03.1531 &  0.7148 &  -19.79 &  -22.28 &  10.72  &     120.3   &   ---     &   4.73   &   ---   & 1.779 & 0.18 & spiral     \\
03.1540 &  0.6898 &  -20.97 &  -22.31 &  10.70  &     123.3   &    115.6  &   4.46   &   ---   & 1.224 & 0.51 & compact   \\
03.1541 &  0.6895 &  -20.09 &    ---  &   ---   &     N/D     &   ---     &   ---    & 8.67    & 1.707 & 0.24 & spiral     \\
03.9003 &  0.6189 &    ---  &    ---  &   ---   &     94.25   &  46.30    &   12.12  &   ---   & 1.622 & 0.30 & irregular \\ \hline 
14.0207 &  0.5460 &  -21.56 &  -23.76 &  11.33  &     N/D     &   ---     &   ---    & ---     & 2.059 & 0.64 & e/s0       \\
14.0302 &  0.5830 &  -20.77 &  -22.87 &  10.93  &     98.03   &   ---     &   ---    & ---     & 1.366 & 0.39 & spiral    \\
14.0393 &  0.6016 &  -21.46 &  -22.30 &  10.67  &     94.97   &   ---     &   12.57  & ---     & 1.490 & 0.22 & spiral    \\
14.0400 &  0.6760 &  -20.72 &  -21.79 &  10.48  &     N/D     &   ---     &   ---    & ---     & 1.437 & 0.35 & spiral       \\
14.0411 &  0.8538 &  -21.33 &    ---  &   ---   &     N/D     &   ---     &   ---    & ---     & 0.748 & 0.43 & compact      \\
14.0422 &  0.4210 &  -20.08 &  -22.29 &  10.73  &     N/D     &   ---     &    0.50  & ---     & 2.078 & 0.31 & e/s0        \\
14.0446 &  0.6030 &  -21.52 &    ---  &   ---   &     89.03   &   ---     &   ---    & ---     & 1.582 & 0.44 & spiral      \\
14.0485 &  0.6545 &  -19.93 &  -21.10 &  10.18  &     N/D     &   ---     &    5.20  & ---     & 1.357 & 0.22 & spiral      \\
14.0547 &  1.1963 &  -22.44 &  -23.83 &  11.28  &     214.8   &   ---     &   ---    & ---     &  ---  &  --- & merger   \\
14.0557 &  0.8800 &  -21.44 &    ---  &   ---   &     N/D     &   ---     &   ---    & ---     & 0.877 & 0.59 & compact     \\\hline  \hline
\end{tabular}
  }
\end{table*}

\setcounter{table}{0}
\begin{table*}
\caption { Cont... }
{\scriptsize
 \begin{tabular}{cccccrrrcccl} \hline
CFRS & z & M$_{B}$  & M$_{K}$ &log(M/M$_{\odot}$)& SFR$_{\rm IR}^a$ &  SFR$_{H\alpha}^a$ &SFR$_{[\rm OII]}^a$ & 12+log(O/H) & C.C.$^b$ & Comp$^c$ & Morph$^d$ \\ \hline
14.0580 &  0.7440 &  -21.07 &    ---  &   ---   &     N/D     &   ---     &   ---    & ---     & 1.849 & 0.28 & spiral      \\
14.0600 &  1.0385 &  -21.75 &    ---  &   ---   &     636.2   &   ---     &   27.75  & ---     &  ---  & ---  & irregular   \\  
14.0651 &  0.5480 &  -19.44 &  -21.75 &  10.51  &     N/D     &   ---     &   ---    & ---     & 1.674 & 0.53 & e/s0        \\  
14.0663 &  0.7434 &  -21.40 &    ---  &   ---   &     187.5   &   ---     &    2.89  & ---     & 1.580 & 0.37 & spiral     \\
14.0665 &  0.8110 &  -19.89 &  -20.85 &  10.07  &     159.5   &   ---     &   ---    & ---     &  ---  & ---  & ---   ---    \\
14.0697 &  0.8271 &  -20.71 &    ---  &   ---   &     N/D     &   ---     &   ---    & ---     & 0.914 & 0.69 & compact     \\
14.0700 &  0.6526 &  -21.29 &    ---  &   ---   &     N/D     &   ---     &   ---    & ---     & 1.905 & 0.65 & e/s0        \\
14.0711 &  1.1180 &  -22.09 &  -23.30 &  11.07  &     N/D     &   ---     &   ---    & ---     & 1.517 & 0.41 & irregular  \\
14.0725 &  0.5820 &  -19.54 &  -21.70 &  10.47  &     41.53   &   ---     &    3.12  & ---     & 1.543 & 0.26 & irregular  \\
14.0727 &  0.4638 &  -20.33 &    ---  &   ---   &     N/D     &   ---     &    7.88  & ---     & 1.804 & 0.35 & spiral      \\ 
14.0746 &  0.6750 &  -20.40 &  -22.64 &  10.88  &     N/D     &   ---     &   ---    & ---     & 1.968 & 0.50 & e/s0         \\
14.0814 &  0.9995 &  -20.50 &  -22.87 &  10.90  &     N/D     &   ---     &   ---    & ---     & 1.636 & 0.17 & irregular  \\
14.0820 &  0.9800 &  -21.53 &  -24.08 &  11.45  &     N/D     &   ---     &   ---    & ---     & 2.300 & 0.43 & e/s0        \\
14.0846 &  0.9900 &  -21.37 &  -22.99 &  10.96  &     N/D     &   ---     &   ---    & ---     & 1.353 & 0.25 & spiral      \\
14.0854 &  0.9920 &  -21.66 &  -23.83 &  11.34  &     N/D     &   ---     &   ---    & ---     & 2.021 & 0.45 & e/s0        \\
14.0899 &  0.8750 &  -21.07 &    ---  &   ---   &     N/D     &   ---     &    4.69  & ---     & 0.438 & 0.34 & spiral      \\
14.0913 &  1.0058 &  -21.58 &    ---  &   ---   &     N/D     &   ---     &   ---    & ---     & 1.946 & 0.58 & e/s0        \\
14.0937 &  1.0097 &  -21.60 &    ---  &   ---   &     N/D     &   ---     &   ---    & ---     & 2.248 & 0.47 & e/s0        \\
14.0962 &  0.7617 &  -21.30 &  -23.02 &  11.02  &     N/D     &   ---     &    4.30  & ---     & 1.900 & 0.52 & spiral      \\
14.0964 &  0.4351 &  -19.32 &    ---  &   ---   &     N/D     &   ---     &   ---    & ---     & 1.232 & 0.27 & spiral      \\
14.0972 &  0.6809 &  -21.10 &  -22.02 &  10.55  &     N/D     &   ---     &   27.06  & ---     & 0.880 & 0.54 & irregular   \\
14.0998 &  0.4300 &  -20.11 &  -22.23 &  10.70  &     44.90   &   ---     &   ---    & ---     & 1.795 & 0.42 & irregular  \\
14.1008 &  0.4362 &  -19.82 &    ---  &   ---   &     N/D     &   ---     &   ---    & ---     & 2.475 & 0.43 & spiral      \\ 
14.1012 &  0.4815 &  -19.85 &  -20.92 &  10.13  &     N/D     &   ---     &    2.36  & ---     & 1.248 & 0.42 & compact     \\      
14.1028 &  0.9876 &  -21.75 &  -23.67 &  11.28  &     N/D     &   ---     &   14.48  & ---     & 2.060 & 0.39 & e/s0         \\
14.1037 &  0.5489 &  -20.31 &  -21.50 &  10.37  &     N/D     &   ---     &    2.45  & ---     & 1.381 & 0.17 & spiral       \\
14.1041 &  0.4351 &  -19.50 &    ---  &   ---   &     N/D     &   ---     &   ---    & ---     &  ---  &  --- & irregular    \\
14.1042 &  0.8916 &  -21.42 &  -23.30 &  11.12  &     161.6   &   ---     &    2.33  & ---     & 1.773 & 0.42 & compact     \\ 
14.1043 &  0.6516 &  -21.60 &  -23.64 &  11.26  &     N/D     &   ---     &  ---   & ---     & 2.035 & 0.42 &  spiral      \\  
14.1079 &  0.9130 &  -21.01 &  -20.95 &  10.15  &     N/D     &   ---     &   8.76 & ---     & 1.633 & 0.33 &  irregular    \\   
14.1087 &  0.6595 &  -20.21 &  -21.08 &  10.17  &     N/D     &   ---     &  5.34  &   ---   & 1.061 & 0.44 &  e/s0         \\
14.1129 &  0.8443 &  -21.19 &  -22.61 &  10.81  &     211.9   &   ---     &  ---   &   ---   &  ---  &  --- &  merger      \\
14.1136 &  0.6404 &  -20.16 &  -19.12 &   9.38  &     N/D     &   ---     & 10.73  &   ---   & 0.723 & 0.43 &  compact      \\
14.1139 &  0.6600 &  -21.60 &  -23.45 &  11.15  &     203.2   &   ---     &  7.70  &   ---   &  ---  &  --- &  merger       \\
14.1146 &  0.7437 &  -20.77 &    ---  &   ---   &     N/D     &   ---     &  8.83  &   ---   & 1.450 & 0.51 &  compact      \\
14.1157 &  1.0106 &  -22.49 &    ---  &   ---   &     N/D     &   ---     &  ---   &   ---   &  ---  &  --- &  merger      \\
14.1164 &  0.6773 &  -20.54 &    ---  &   ---   &     N/D     &   ---     &  ---   &   ---   & 1.066 & 0.30 &  irregular    \\
14.1179 &  0.4345 &  -19.13 &    ---  &   ---   &     N/D     &   ---     &  ---   &   ---   & 1.953 & 0.50 &  e/s0        \\
14.1189 &  0.7526 &  -20.42 &  -21.57 &  10.39  &     N/D     &   ---     &  5.33  &   ---   & 1.143 & 0.35 &  spiral       \\
14.1190 &  0.7544 &  -21.38 &    ---  &   ---   &     159.1   &   ---     &  3.06  &   ---   & 1.583 & 0.35 &  spiral       \\
14.1213 &  0.7631 &  -21.03 &    ---  &   ---   &     N/D     &   ---     &  ---   &   ---   &  ---  &  --- &  merger      \\
14.1223 &  0.6936 &  -20.42 &    ---  &   ---   &     N/D     &   ---     &  ---   &   ---   & 1.078 & 0.32 &  spiral      \\
14.1232 &  0.7660 &  -20.22 &  -22.73 &  10.85  &     N/D     &   ---     &  ---   &   ---   &  ---  &  --- &  merger      \\
14.1246 &  0.6486 &  -19.76 &    ---  &   ---   &     N/D     &   ---     &  ---   &   ---   & 2.140 & 0.27 &  irregular   \\
14.1277 &  0.8100 &  -21.33 &    ---  &   ---   &     N/D     &   ---     &  6.91  &   ---   & 1.917 & 0.52 &  spiral      \\
14.1305 &  0.8069 &  -20.59 &    ---  &   ---   &     N/D     &   ---     &  ---   &   ---   &  ---  &  --- &  merger   \\
14.1311 &  0.8065 &  -22.03 &    ---  &   ---   &     N/D     &   ---     &  ---   &   ---   & 2.022 & 0.52 &  e/s0      \\
14.1326 &  0.8111 &  -20.74 &    ---  &   ---   &     N/D     &   ---     &  ---   &   ---   & 1.200 & 0.20 &  irregular  \\
14.1329 &  0.3750 &  -20.63 &    ---  &   ---   &     22.43   &   ---     &  3.48  &   ---   &  ---  &  --- &   ---            \\
14.1350 &  1.0054 &  -21.29 &    ---  &   ---   &     N/D     &   ---     &  ---   &   ---   & 1.631 & 0.18 &  irregular  \\
14.1355 &  0.4801 &  -19.74 &  -20.63 &  10.01  &     N/D     &   ---     &  3.94  &   ---   & 1.412 & 0.26 &  spiral      \\
14.1356 &  0.8307 &  -20.44 &    ---  &   ---   &     N/D     &   ---     &  7.61  &   ---   & 1.739 & 0.42 &  compact   \\
14.1386 &  0.7410 &  -21.22 &  -22.72 &  10.84  &    132.32   &   ---     &  8.50  &   ---   &  ---  & ---  &   ---            \\
14.1395 &  0.5301 &  -19.85 &    ---  &   ---   &     N/D     &   ---     &  4.47  &   ---   & 1.070 & 0.27 &  spiral      \\
14.1412 &  0.9122 &  -21.03 &    ---  &   ---   &     N/D     &   ---     &  ---   &   ---   & 1.729 & 0.25 &  spiral      \\
14.1415 &  0.7486 &  -21.15 &    ---  &   ---   &     N/D     &   ---     &  ---   &   ---   & 2.014 & 0.48 &  compact    \\
14.1427 &  0.5379 &  -19.77 &    ---  &   ---   &     N/D     &   ---     &  ---   &   ---   & 1.459 & 0.42 &  spiral      \\
14.1444 &  0.7420 &  -19.89 &  -22.57 &  10.82  &    139.57   &   ---     &  2.91  &   ---   &  ---  & ---  &   ---            \\
14.1464 &  0.4620 &  -19.62 &    ---  &   ---   &     N/D     &   ---     &  0.69  &   ---   & 2.014 & 0.56 &  e/s0        \\
14.1477 &  0.8191 &  -21.02 &    ---  &   ---   &     N/D     &   ---     &  ---   &   ---   & 2.002 & 0.52 &  e/s0        \\
14.1496 &  0.9163 &  -21.06 &    ---  &   ---   &     N/D     &   ---     & 14.01  &   ---   & 1.097 & 0.55 &  compact      \\
14.1501 &  1.0018 &  -21.41 &    ---  &   ---   &     N/D     &   ---     &  ---   &   ---   &  ---  &  --- &  irregular   \\
14.1506 &  1.0126 &  -21.51 &    ---  &   ---   &     N/D     &   ---     &  ---   &   ---   & 1.942 & 0.59 &  e/s0        \\
14.1524 &  0.4297 &  -21.16 &    ---  &   ---   &     N/D     &   ---     &  ---   &   ---   & 1.363 & 0.31 &  spiral      \\
14.1525 &  0.7480 &  -20.62 &  -22.00 &  10.55  &    250.90   &   ---     & 13.10  &   ---   &  ---  & ---  &   ---            \\ \hline \hline 
\end{tabular}
  }
\end{table*}

\setcounter{table}{0}
\begin{table*}
\caption { Cont... }
{\scriptsize
 \begin{tabular}{cccccrrrcccl} \hline
CFRS & z & M$_{B}$  & M$_{K}$ &log(M/M$_{\odot}$)& SFR$_{\rm IR}^a$ &  SFR$_{H\alpha}^a$ &SFR$_{[\rm OII]}^a$ & 12+log(O/H) & C.C.$^b$ & Comp$^c$ & Morph$^d$ \\ \hline
14.1554 &  0.9115 &  -21.04 &    ---  &   ---   &     N/D     &   ---     &  ---   &   ---   & 1.388 & 0.41 &  spiral      \\
14.1567 &  0.4800 &  -21.83 &  -22.91 &  10.88  &     78.20   &   ---     &  ---   &   ---   &  ---  & ---  &   ---            \\
14.1601 &  0.5370 &  -20.36 &    ---  &   ---   &     N/D     &   ---     &  ---   &   ---   & 1.227 & 0.34 &  spiral      \\
14.1620 &  0.9182 &  -21.53 &    ---  &   ---   &     N/D     &   ---     &  ---   &   ---   & 2.081 & 0.39 &  e/s0        \\ \hline  
22.0242 &  0.8638 &  -20.53 &  -22.27 &  10.68  &     ---     &   151.0   &  6.25  & 8.52    &  ---  &  --- &  ---            \\
22.0344 &  0.5168 &  -19.79 &  -19.84 &   9.70  &     ---     &   33.89   & 10.90  & 8.52    &  ---  &  --- &  ---     \\
22.0355 &  0.4317 &  -20.23 &    ---  &   ---   &     ---     &   ---     &  ---   & 8.87    &  ---  &  --- &  ---     \\
22.0429 &  0.6243 &  -19.93 &  -21.72 &  10.48  &     ---     &   50.20   &  2.48  & 8.61    &  ---  &  --- &  ---     \\ 
22.0500 &  0.5106 &  -19.92 &    ---  &   ---   &     ---     &   46.73   &  ---   & 8.27    &  ---  &  --- &  ---     \\     
22.0576 &  0.8905 &  -20.82 &    ---  &   ---   &     ---     &   36.44   & 19.21  & 8.58    &  ---  &  --- &  ---     \\
22.0599 &  0.8854 &  -21.22 &    ---  &   ---   &     ---     &   161.7   & 32.26  & 8.32    &  ---  &  --- &  ---     \\ 
22.0619 &  0.4671 &  -19.21 &    ---  &   ---   &     ---     &   5.210   &  ---   & 8.36    &  ---  &  --- &  ---      \\
22.0626 &  0.5150 &  -19.36 &  -19.64 &   9.59  &     ---     &   1.998   &  ---   & 8.47    &  ---  &  --- &  ---      \\
22.0637 &  0.5419 &  -20.66 &  -21.37 &  10.32  &     ---     &   10.36   & 11.35  & 8.55    &  ---  &  --- &  ---      \\       
22.0721 &  0.4070 &  -18.83 &  -20.46 &   9.99  &     ---     &   13.89   &  ---   &   ---   &  ---  &  --- &  ---      \\
22.0779 &  0.9252 &  -21.08 &  -22.37 &  10.70  &     ---     &   31.66   &  6.38  & 8.58    &  ---  &  --- &  ---      \\
22.0828 &  0.4070 &  -18.75 &  -20.53 &  10.01  &     ---     &   30.20   &  1.25  &   ---   &  ---  &  --- &  ---     \\
22.0919 &  0.4714 &  -19.92 &    ---  &   ---   &     ---     &   7.182   &  0.83  & 8.17    &  ---  &  --- &  ---     \\
22.1064 &  0.5369 &  -19.76 &  -21.64 &  10.40  &     ---     &   24.61   &  5.65  & 8.46    &  ---  &  --- &  ---         \\
22.1070 &  0.8796 &  -20.49 &    ---  &   ---   &     ---     &   107.6   &  ---   & 8.55    &  ---  &  --- &  ---     \\     \hline
UDSR04  & 0.4501 &   -19.35  &   ---   &  ---    &   15.68     &   ---     &  2.21  &   ---   &  ---  &  --- &  --- \\
  05    & 1.0660 &   -22.18* &   ---   &  ---    &   183.1     &   ---     & 19.92  &   ---   &  ---  &  --- &  --- \\
  08    & 0.7291 &   -21.23  &   ---   &  ---    &   35.67     &    16.05  & 16.88  &   8.66  &  ---  &  --- &  --- \\
  10    & 0.6798 &   -22.09  &   ---   &  ---    &   96.92     &    42.08  &  7.21  &   8.71  &  ---  &  --- &  --- \\
  14    & 0.8150 &   -21.98  &   ---   &  ---    &   38.00     &    84.41  & 18.38  &   8.78  &  ---  &  --- &  --- \\
  15    & 0.4949 &   -20.06  &   ---   &  ---    &   18.40     &    37.77  &  ---   &   8.81  &  ---  &  --- &  --- \\
  17    & 0.9680 &   -22.08  &   ---   &  ---    &   219.0     &   ---     &  4.64  &   ---   &  ---  &  --- &  --- \\
  20    & 0.7660 &   -21.87* &   ---   &  ---    &   35.81     &    84.95  & 26.64  &   8.65  &  ---  &  --- &  --- \\
  21    & 0.4655 &   -18.68  &   ---   &  ---    &   3.01      &   ---     &  0.90  &   ---   &  ---  &  --- &  --- \\
  23    & 0.7094 &   -21.86  &   ---   &  ---    &   40.96     &    74.07  &  7.38  &   8.72  &  ---  &  --- &  --- \\
  26    & 0.3841 &   -19.67  &   ---   &  ---    &   3.17      &   ---     &  1.01  &   8.44  &  ---  &  --- &  --- \\
  27    & 0.6429 &   -20.67  &   ---   &  ---    &   10.93     &   ---     &  1.81  &   ---   &  ---  &  --- &  --- \\
  30    & 0.9590 &   -21.51  &   ---   &  ---    &   225.9     &   ---     &  3.41  &   ---   &  ---  &  --- &  --- \\
  32    & 0.5841 &   -19.98* &   ---   &  ---    &   10.13     &    14.98  &  6.23  &   8.42  &  ---  &  --- &  --- \\  \hline
UDSF01  & 0.4656 &   -21.23  &   ---   &  ---    &   4.82      &    11.26  & 10.19  &   8.63  &  ---  &  --- &  --- \\
  02    & 0.7781 &   -20.84  &   ---   &  ---    &   4.60      &   ---     &  4.14  &   8.56  &  ---  &  --- &  --- \\
  03    & 0.5532 &   -20.52  &   ---   &  ---    &   11.14     &   ---     &  ---   &   8.82  &  ---  &  --- &  --- \\
  04    & 0.9620 &   -21.55  &   ---   &  ---    &   97.17     &   ---     &  3.17  &   ---   &  ---  &  --- &  --- \\
  06    & 0.6928 &   -21.49  &   ---   &  ---    &   15.97     &   ---     &  3.46  &   ---   &  ---  &  --- &  --- \\
  07    & 0.7014 &   -20.70  &   ---   &  ---    &   16.77     &    11.43  &  2.86  &   8.81  &  ---  &  --- &  --- \\ 
  08    & 0.7075 &   -21.03* &   ---   &  ---    &   14.01     &   ---     &  2.19  &   ---   &  ---  &  --- &  --- \\
  12    & 0.7388 &   -20.95  &   ---   &  ---    &   62.83     &   ---     &  5.99  &   ---   &  ---  &  --- &  --- \\
  13    & 0.7605 &   -20.07  &   ---   &  ---    &   408.5     &   ---     &  2.71  &   ---   &  ---  &  --- &  --- \\
  14    & 0.8190 &   -21.48  &   ---   &  ---    &   7.70      &   ---     &  2.51  &   ---   &  ---  &  --- &  --- \\
  16    & 0.4548 &   -21.42  &   ---   &  ---    &   8.16      &    31.65  &  1.70  &   8.93  &  ---  &  --- &  --- \\
  17    & 0.8100 &   -22.42* &   ---   &  ---    &   56.97     &    38.18  & 11.47  &   8.56  &  ---  &  --- &  --- \\
  18    & 0.4620 &   -20.60* &   ---   &  ---    &   11.03     &     6.20  &  4.77  &   8.66  &  ---  &  --- &  --- \\
  19    & 0.5476 &   -21.94* &   ---   &  ---    &   80.86     &    60.94  &  4.71  &   8.92  &  ---  &  --- &  --- \\
  21    & 0.6980 &   -21.35* &   ---   &  ---    &   16.20     &    14.40  &  ---   &   8.74  &  ---  &  --- &  --- \\
  24    & 1.1590 &   -23.31* &   ---   &  ---    &   138.9     &   ---     & 57.02  &   ---   &  ---  &  --- &  --- \\
  25    & 0.8094 &   -20.42* &   ---   &  ---    &   12.52     &   ---     & 11.46  &   ---   &  ---  &  --- &  --- \\ 
  26a   & 0.7023 &   -21.60* &   ---   &  ---    &   16.31     &    10.45  & 12.68  &   8.36  &  ---  &  --- &  ---   \\
  26b   & 0.7027 &   -20.62* &   ---   &  ---    &   18.38     &    27.11  &  ---   &   8.62  &  ---  &  --- &  ---  \\
  28    & 0.6621 &   -21.84* &   ---   &  ---    &   42.96     &   ---     &  ---   &   ---   &  ---  &  --- &  --- \\
  29    & 0.6619 &   -20.45* &   ---   &  ---    &   51.34     &    23.34  &  2.44  &   8.43  &  ---  &  --- &  --- \\
  31    & 0.6868 &   -20.57* &   ---   &  ---    &   21.71     &   ---     &  1.56  &   8.82  &  ---  &  --- &  --- \\
  32    & 0.7268 &   -19.73* &   ---   &  ---    &   32.28     &   ---     &  1.25  &   ---   &  ---  &  --- &  --- \\    \hline \hline
\end{tabular}
  }
\begin{list}{}{}
\item[ ---] Not available
\item[N/D] Not detected 
\item[$^{\mathrm{a}}$] Star formation rate in M$_{\odot}$/yr.
\item[$^{\mathrm{c, d \; \&\; e }}$] Central color, compactness and morphology  from HST images
\item[$^{*}$] $M_{B}$ estimated from observed R band and from the observed spectroscopic slope 
\end{list}

\end{table*}

\section{Towards more robust estimates of star formation in distant galaxies}

A prerequisite for estimating star formation from emission lines is to properly estimate the extinction. $H\alpha$ luminosity is one of the best indicators of the instantaneous SFR.  However, low resolution spectroscopy (R $< 500$) often produces misleading results (Liang et al. 2004a). Only good $S/N$ spectroscopy with moderate spectral resolution (R$>600$) allows a proper estimate of the extinction from the $H\alpha/H\beta$ ratio after accounting for underlying stellar absorption. SFRs can be otherwise underestimated  or overestimated by factors reaching 10, even if one accounts for an {\it ad hoc} extinction correction. These effects are prominent for a large fraction of evolved massive galaxies especially those experiencing successive bursts (A and F stars dominating their absorption spectra).  

ISOCAM observations have provided a unique opportunity to test the validity of our SFR estimates. It has been shown (Elbaz et al. 2002) that the
mid-IR and radio luminosities correlate well up to $z=1$. Bolometric IR measurements derived from the mid-IR are validated, unless if both the radio-FIR and the MIR-FIR correlations have strongly changed at high redshifts. Mid-IR photometry provides a unique tool to estimate properly the SFR of LIRGs (defined as $L_{IR} >$  $10^{11}$ $L_{\odot}$) up to $z=1.2$, and of starbursts ($L_{IR} <$  $10^{11}$ $L_{\odot}$) up to $z=0.4$. For a sample of 16 ISO galaxies at $0< z < 1$,  SFRs have been derived from the extinction-corrected $H\alpha$ luminosities (Flores et al. 2004). These values agree within a factor 2 with mid-IR estimates (median value of $SFR_{IR}/SFR_{H\alpha}$=1.05). Using moderate spectral resolution (R$= 1200$) with a good $S/N$ of 90 ISO galaxies ($0.2< z < 1$), Liang et al. (2004b) have shown that the extinction can be properly derived from the $H\beta/H\gamma$ ratio, using high S/N spectroscopy at moderate resolution at VLT. Such measurements can be performed for all galaxies up to $z=1$, and Liang et al (2004b) find that the derived SFR from extinction corrected $H\beta$ line is consistent with $SFR_{IR}$, with their ratio ranging from 0.6 to 1.7. 

Can we also use [OII]$3727$ luminosity to derive SFRs of high redshift galaxies using spectrographs in the visible range ? 
Figure 1 shows that it systematically underestimates the SFR by large factors,  reaching  median values of 5 to 22 for starbursts and for LIRGs, respectively.
 The fact that UV fluxes provide even lower SFR values than [OII]$3727$ fluxes might be related to the expected increase of the [OII]$3727$/$H\beta$ ratio with decreasing metal abundance (high z systems being expected to be less metal abundant than local ones).  An important word of caution is however required here, since using the UV slope, Meurer, Heckman \& Calzetti (1999) were able to derive extinction corrections after comparison to modelling. By comparing these corrections to IR fluxes of z$\sim$ 0.9 ISO galaxies from Flores et al. (1999), Adelberger \& Steidel (2000) claimed that their method appears to work within a factor 3 (see also Adelberger, 2001). On the other hand, Bell (2002) argues that the rest-frame UV and optical data provides estimates of the star formation rates which are subject to large systematic uncertainties because of dust, which cannot be reliably corrected for using only UV/optical diagnostics.    
 
In summary, the star formation for strong star forming systems is severely underestimated by using UV tracers alone, which is not surprising given the expected effects of dust extinction. It is hardly a new result since the discovery of LIRGs \& ULIRGs by IRAS. We will see in the following that, unlike the situation in the local Universe, such an underestimation has profound implications for our understanding of the evolution in the distant Universe.

\begin{figure*}[]
\begin{center}
\includegraphics[width=1.0\textwidth]{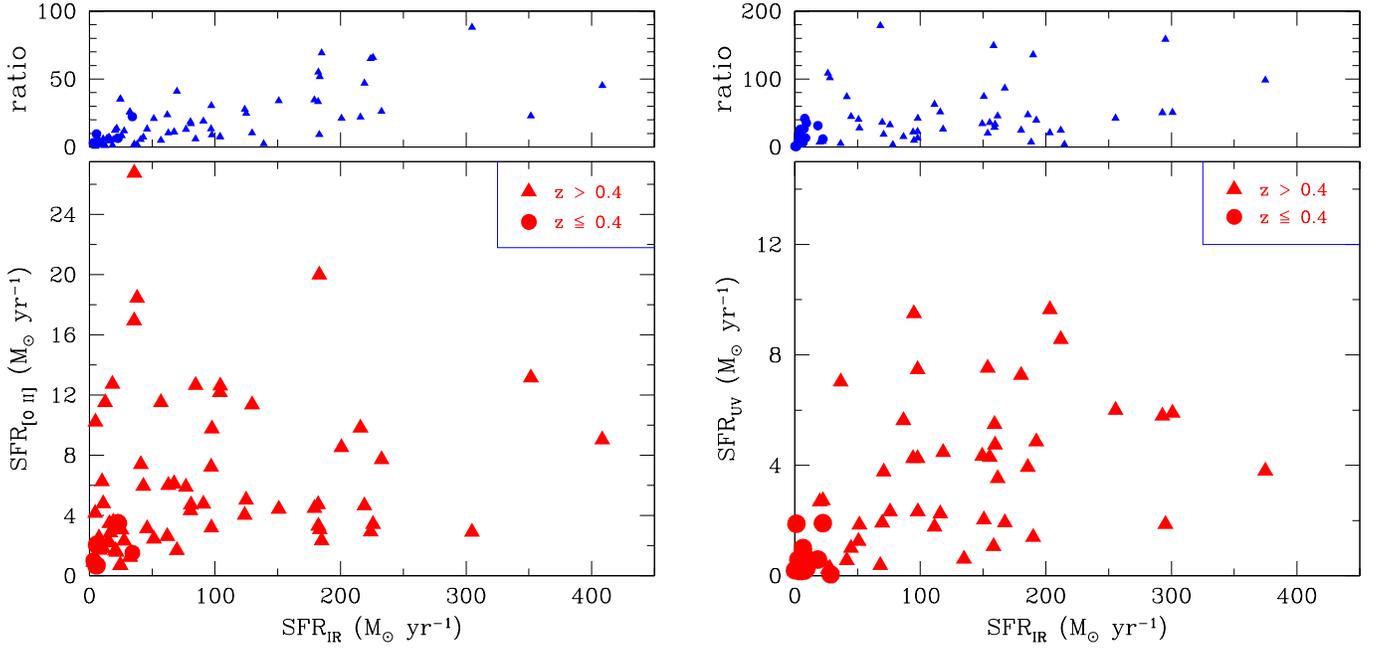}
\end{center}
\caption[]{ Comparison of SFRs calculated using the formalism of Kennicutt (1998).
({\it Left panel}): [OII]$\lambda$3727 estimates are compared to IR estimates for a sample of 70
galaxies from the CFRS and Marano fields (see Liang et al. 2004b). Upper left panel provides
the $SFR_{IR}/SFR_{[OII]}$ ratio. Median values of the ratio are 5 and 22 for starbursts (SFR $<$
20 $M_{\odot} yr^{-1}$) and LIRGs respectively. ({\it Right panel}): same comparison for
2800\AA~ UV luminosity estimated for 61 CFRS galaxies. Median values of the 
$SFR_{IR}/SFR_{UV}$ ratio are 13 and 36 for starbursts and LIRGs, respectively.}
\label{eps1}
\end{figure*}

\begin{figure*}
\begin{center}
\includegraphics[width=0.8\textwidth]{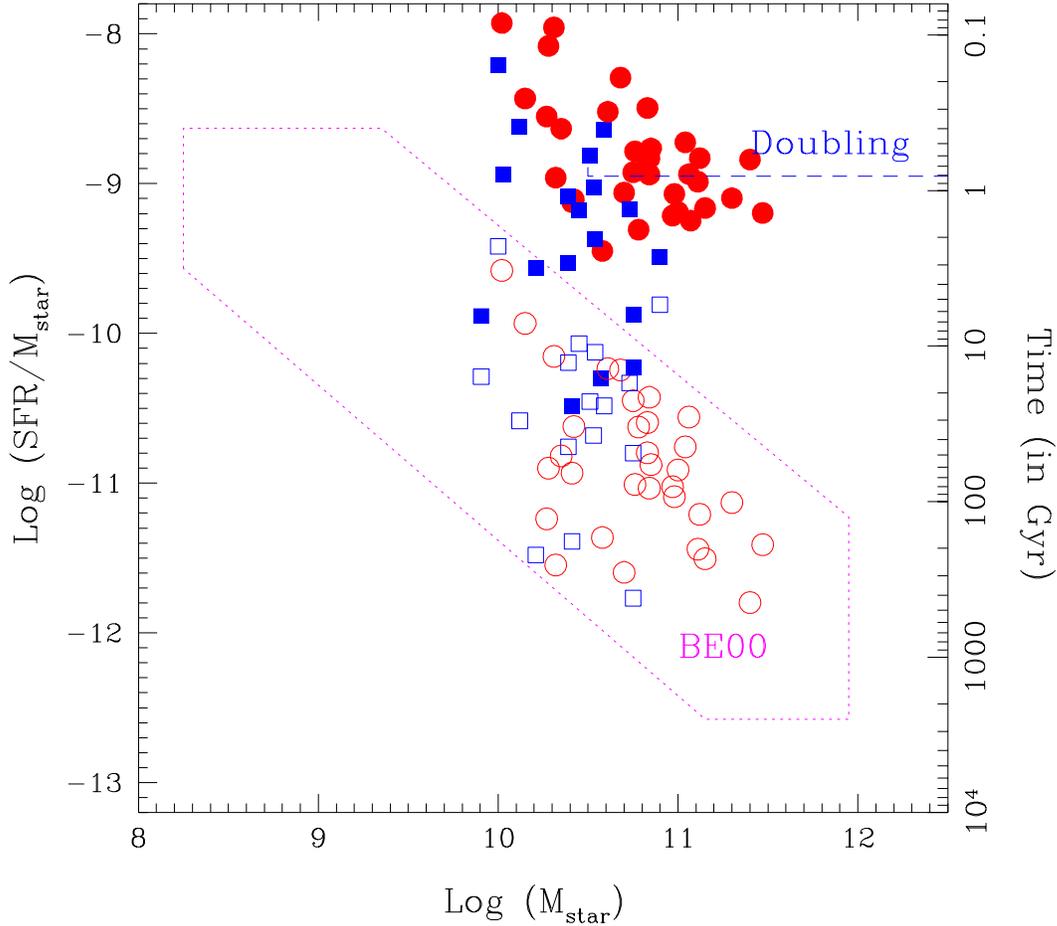}
\end{center}
\caption[]{  Typical timescale for forming stellar mass in $z>0.4$ LIRGs (full dots) and LCGs (filled squares, no LIRGs included) related to their observed stellar masses. SFRs have been estimated from extinction-corrected Balmer emission lines (LCGs) or IR luminosities (LIRGs).  The dotted line box indicates the position occupied by the galaxies when their SFR is derived from [OII]3727 line observations, following the adopted calibration from Brinchmann and Ellis (2000). Similar estimates have been made for this sample for LIRGs (empty dots) and for compact galaxies (empty squares). This shows that an important part of the star formation can be hidden using UV estimates of the SFR. The horizontal dashed line shows the median doubling time-scale for the population of intermediate mass LIRGs. There is a hint that LIRG masses could have been overestimated by a factor of $\sim$ 2 (see section 7); this would move the full dots towards the upper-left, leading to a median characteristic time of 0.4 Gyr, while the total mass formed would be unchanged (SFR still stands). Accurate determinations of SFR/M for LIRGs will be provided by Marcillac et al. (2004, in preparation)   }
\label{eps2}
\end{figure*}

\begin{figure*}
\begin{center}
\includegraphics[width=0.8\textwidth]{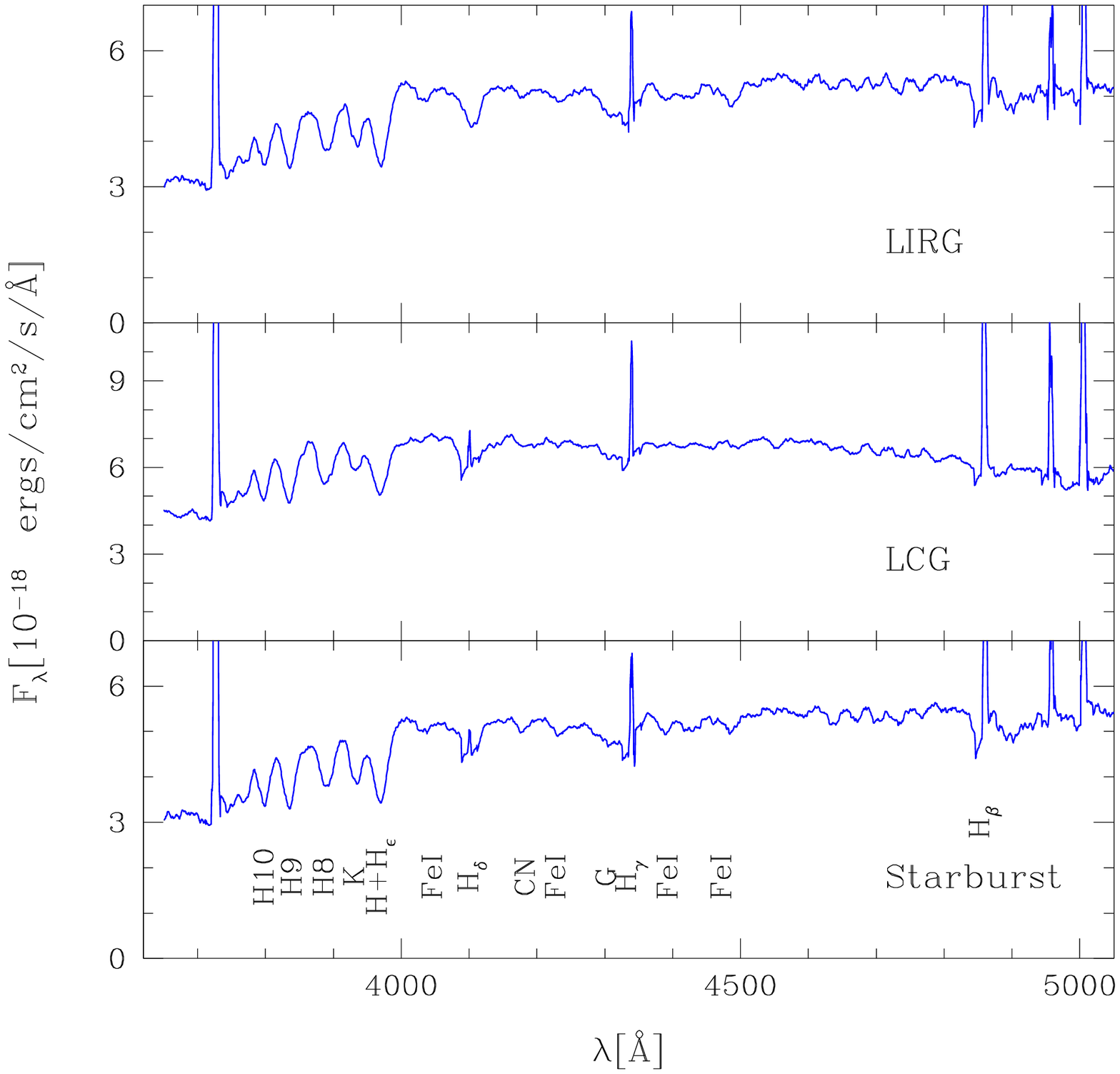}
\end{center}
\caption[]{Median combined VLT/FORS (R600+I600) spectra of 38 distant LIRGs, 21 LCGs (LIRGs have been excluded here) and 26 normal starbursts showing their absorption lines. A 3$\sigma$ clipping method has been used to reject residual unrelated features (mostly from sky substraction). All spectra invariably show a mixture of intermediate age stellar population revealed by their strong Balmer absorption system with an older stellar population revealed by metallic absorption lines.}
\label{eps3}
\end{figure*}

\section{How and when did galaxies form the bulk of their stars and metals ?}

\subsection{Luminous IR galaxies: a rare phenomenon locally, a common one at $z > 0.4$}
LIRGs are rare objects in the local Universe (Sanders and Mirabel, 1996). Soifer et al. (1986) estimated that 0.5\% of galaxies with visible luminosities larger than $10^{10} L_{\odot}$, have spectral energy distribution typical of LIRGs (90\% of their luminosity emitted in IR). How this compares to the distant Universe ?

A total of 113 galaxies have been detected at 15$\mu$m by ISOCAM (Flores et al. 1999, 2004) in the CFRS 14h and 03h fields (2 x 100 $arcmin^{2}$), and 70\% of them are LIRGs at $z > 0.4$, all with $M_{B}\le -$20. Compared with the 311 $z > 0.4$ CFRS galaxies occupying a solid angle of 112 $arcmin^{2}$ (Hammer et al. 1997), it appears that 15\% of the $M_{B}\le -$20 galaxies at $z> 0.4$ are indeed LIRGs. A similar calculation can be made by considering the subsample of 184 galaxies possessing K band photometry, providing a similar fraction (15\%) of LIRGs in M $>$ 5 $10^{10}$ M$_{\odot}$ galaxies at $z >0.4$. This fraction is remarkably stable to varying the lower mass limit from 3 $10^{10}$ M$_{\odot}$ to 10 $10^{10}$ M$_{\odot}$. The marginal incompleteness of the CFRS (85\% of  redshift identified) can slightly decrease the LIRG fraction and it is mostly related to the difficulty in deriving a redshift for distant red galaxies near the CFRS $I_{AB}$ limit (Crampton et al. 1995; Hammer et al. 1997). On the other hand, the fraction of LIRGs is underestimated because the depth of the ISO observations of the CFRS fields (complete at S(15$\mu$m) $>$ 300 $\mu$Jy) does not ensure that all LIRGs near the cut-off limits (L$_{IR}=10^{11} L_{\odot}$) have been detected, especially at the high-z end (see the discussion by Liang et al. 2004b). A gross consistency check is provided by comparing the mid-IR counts from Elbaz et al. (1999) to those from optical, knowing that the redshift distribution of mid-IR faint sources is roughly comparable to that of the CFRS. The number density of 100$\mu$Jy sources is 31\% that of $I_{AB}<$ 22.5, and the Franceschini et al. (2003)' study of such sources in the HDF South find that 11/21 of them are indeed LIRGs with $I_{AB}<$ 22.5. This is independent support that at z$>$ 0.4, 15\% of the CFRS galaxies are LIRGs.

\subsection{Stellar mass formed during the time between $z=1$ and $z=0.4$}
The ratio of stellar mass to SFR provides the time-scale to double the stellar mass content, assuming a constant SFR.  Figure 2 relates the doubling timescale to the stellar masses of LIRGs and LCGs. Re-evaluating the SFRs of galaxies from IR data leads to doubling time-scales from 0.1 to 1.1 Gyr for LIRGs, i.e. much shorter than those derived in earlier work (BE2000). Here we investigate how the new stars formed in LIRGs can impact the stellar mass of the galaxy population in the intermediate redshift range. 

 The intermediate stellar mass galaxies ($10.5 < log(M/M_\odot)< 11.5$) include 15\% of LIRGs with a median doubling time-scale of 0.8 Gyr (Figure 2), in excellent agreement with the median value from Franceschini et al. (2003, see their Figure 10) for S(15$\mu$m) $>$ 100 $\mu$Jy galaxies within the same mass range. The newly formed stellar mass in LIRGs during the elapsed time of 3.3 Gyr (from $z=1$ to $z=0.4$) correspond to $0.15 \times 3.3/0.8 = 62$\% of the $z=1$ total mass of intermediate mass galaxies. Flores et al. (1999) found that by including the IR contribution, the CSFD is between 2 to 3 times larger than that estimated from UV. Including all kind of star formation (IR \& UV emittors), it is expected that the recent star formation  produces from 70\% to 100\% of the total mass of intermediate mass galaxies at $z=1$ (or alternatively 40 to 50\% of the mass at $z=0$).

The star formation in LIRGs provides a substantial stellar mass production in intermediate mass galaxies, because of their large number density at high redshifts, in constrast with the situation in the local Universe. The extremely tiny fraction of recently formed mass by BE2000 is due to their use of the [OII]$3727$ line as a SFR tracer, and they indeed commented that this choice could lead to serious underestimates. Our result is however in good agreement with methods independent of SFR estimates from UV tracers, such as those based on fossil records of present-day intermediate mass galaxies (H2004). Integrations from $z=1$ to $z=0.4$ of their SFR history lead to a stellar mass increase by $\sim$110\% of their total mass at $z=1$. A non-negligible mass accretion from satellites (see Bundy et al. 2004) could also contribute to the stellar mass assembly in intermediate mass galaxies. 
  
\subsection{Successive infra-red episodes (IREs): a plausible star formation history for individual galaxies}

How can we interpret such a large number density of LIRGs in the distant Universe ? Let us examine an extreme hypothesis that LIRGs are a specific galaxy population, and are actively forming stars at very high rate during the 3.3 Gyr from $z=1$ to $z=0.4$. During that time these galaxies would multiply their masses by $ 2 \times (3.3/0.8)=8.2$, producing galaxies in the mass range $11.4 < log(M/M_\odot) < 12.4$, which is dominated by massive ellipticals (Nakamura et al. 2004). Besides the theoretical challenge in interpreting a sustained star formation at such high rates ($\sim 50$M$_\odot/yr$), recent formation of most of their stars in a noticeable fraction of massive ellipticals in the field has never been detected by studies of their local properties. It is generally argued that they are  formed at a much higher redshifts, possibly by monolithic collapses (Renzini, 1999). 

The only plausible alternative is that LIRGs are related to the population of intermediate mass galaxies, probably through an episodic star formation history. Hierarchical models in the $\Lambda$CDM cosmogony predict that galaxies formed through a number of mergers leading to star formation histories with multiple short-living peaks (Abadi et al. 2003). A significant fraction of LIRGs are indeed associated with mergers at different stages after careful examinations of their morphologies (Flores et al. 1999; Aussel et al. 1999; Zheng et al. 2004a). Moreover, distant LIRGs display optical properties astonishingly similar to those of other distant field galaxies. Optical properties are unable to properly distinguish LIRGs from the more quiescent field population. 
As for other distant galaxies, their spectra (Figure 3) invariably show the combination of an old stellar population, with strong Balmer absorption systems (A+F stars) and emission lines.

\begin{figure*}
\begin{center}
\includegraphics[width=0.7\textwidth]{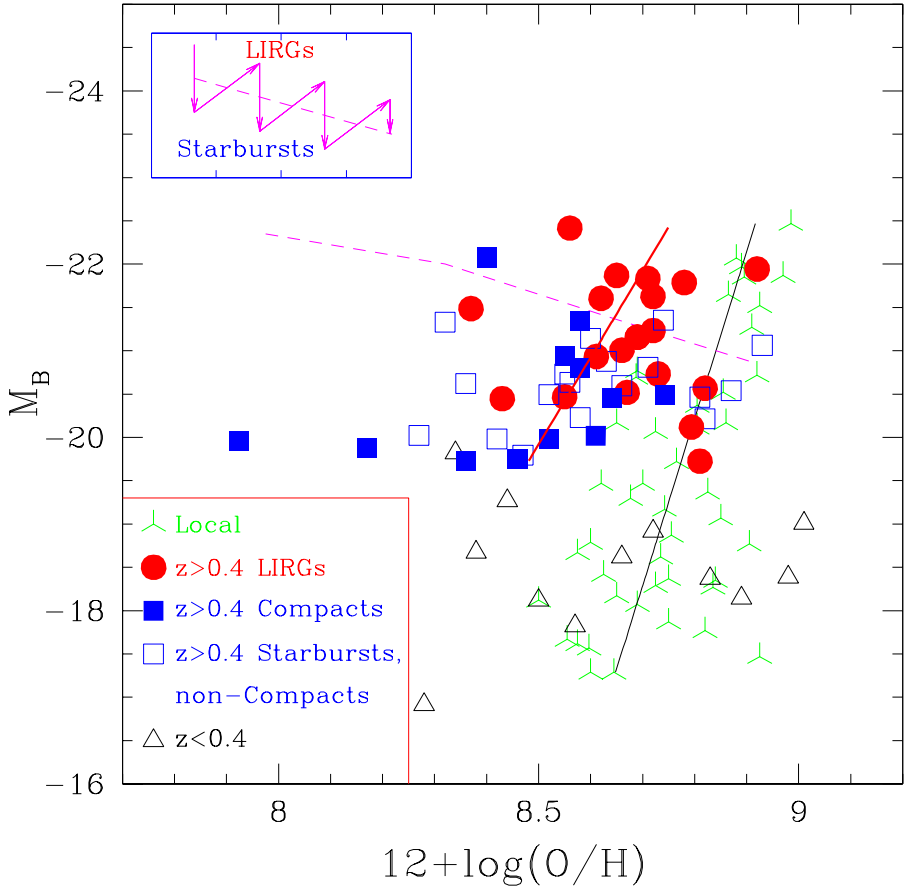}
\end{center}
\caption[]{ M$_B$ versus $O/H$ for CFRS galaxies (16 LIRGs, 12 LCGs and 19 starbursts at z$>$ 0.4). Most low-z sources ($z< 0.4$) lie near the local spiral sequence, in contrast with $z> 0.4$ galaxies which on average, are metal deficient by a factor of 2 (see text). Data for local spirals and for z$<$ 0.4 galaxies are described in Liang et al (2004b).
}
\label{eps4}
\end{figure*}

Therefore, the most plausible explanation of the high fraction of LIRGs in the distant Universe is that they correspond to episodes of violent star formation that are superimposed on relatively low levels of (secular?) star formation activity in more quiescent galaxies. It is generally believed that typical  timescales for strongly enhanced star formation in LIRGs are $\sim 0.1$Gyr (Franceschini et al. 2003), and this is consistent with peaks of star formation occuring during merging events described by hydrodynamical models (Tissera et al. 2002; Scannapieco \& Tissera, 2003). 

If we assume that all intermediate mass galaxies could potentially experience an infrared episode (IRE), during a characteristic time $\tau_{\rm IRE}$, then the number of episodes per galaxy is $n_{\rm IRE}= 0.15 \times$ ($3.3Gyr/ \tau_{\rm IRE})= 5 (\tau_{\rm IRE}/0.1Gyr)^{-1}$. A lower limit on $\tau_{\rm IRE}$ is provided by the lifetime of massive stars and dust formation (typically 0.01 Gyr). Enhanced star formation during merging episodes could have durations of several tenths of Gyr (see a description in Tissera et al. 2002). An important support for $\tau_{\rm IRE}\sim$0.1 Gyr is provided by Marcillac et al. (2004, in preparation) who can reproduce the Balmer absorption line H8 and the 4000$\AA$ break of LIRGs using a large set of star formation histories from Monte Carlo simulations based on Bruzual and Charlot (2004) models. Several infrared episodes per galaxy from $z=1$ and $z=0.4$ can easily account for the large number of $z> 0.4$ CFRS galaxies which show an important population of intermediate age stars (Hammer et al. 1997).   

\subsection{The luminosity-metal abundance relation}

The gas phase metal content of 16 distant LIRGs and 12 LCGs  was derived from VLT spectra (Hammer et al. 2001; Liang et al. 2004b; Gruel, 2002). The same methodology was applied (see Liang et al. 2004b for a detailed description) to evaluate the $O/H$ ratio from emission lines of 19 additional $z>0.4$ normal starbursts, which form fewer stars than LIRGs and are more extended than LCGs. Figure 4 shows that all exhibit a very broad distribution of their gas abundances. There is a hint that LIRGs show slightly higher metal abundances than the rest of the sample, which is not unexpected given the high production of metals in such strong star formation sites. Such an expectation might however be blurred by the 0.08 dex typical error bars (Liang et al. 2004b). 

On average $z>0.4$ galaxies show metal abundances two times lower than those of the local spiral sequence (Figure 4). If, as we propose later (see section 6), most intermediate mass galaxies at $z > 0.4$ are connected to present-day spirals, this provides  independent support that they (at least those with emission lines) have been actively forming about half of their present-day stellar and metal content from $z\sim$ 1 to 0.4. All trends in Figure 4 can be easily accounted for by episodic star formation (displayed by LIRGs): the broad range of metal abundance of LIRGs takes its origin from that of the underlying population of emission line galaxies. The dashed line in Figure 4 represents a simple infall model, assuming a 1 Gyr infall time and a total mass of 10$^{11}$ M$_\odot$ (see Kobulnicky et al. 2003). A more complex evolutionary track probably applies to distant emission line galaxies, as illustrated in the top left of the figure. After a LIRG episode the luminosity decreases because of the disappearance of short-lived massive blue stars. During a LIRG episode the ISM is enriched and the luminosity increases due to the burst. The amplitude of both fading and burst brightening is however attenuated by the dust produced by massive stars. Dust extinction in LIRGs is severe enough that the UV luminosity of LIRGs is, on average, only $\sim$2 times greater than that of starbursts (see Figure 1),  which is consistent with their relative M$_B$ distributions.

Keck and VLT studies (Kobulnicky et al. 2003; Maier, Meisenheimer and Hippelein, 2004) find a similar metal deficiency for $z>0.4$ galaxies as faint as M$_B=-18$. Taken together, these results indicate that, as an ensemble, emission line galaxies were poorer in metals at $z \sim 0.7$ than present day spirals by a factor of order two.

\begin{table}
\caption{Morphological classification statistics for 36  LIRGs and 75 non LIRGs ($z>0.4$), based on state of the art automatic classification schemes (Zheng et al. 2004a; also Zheng et al. 2004b); for comparison, the last column shows the fractions derived from Nakamura et al (2004) for the same galaxy mass range. }
\begin{tabular}{lcccc} \hline
Type   & $z>0.4$ & $z>0.4$  & $z>0.4$ & local \\ 
       & LIRGs & non-LIRGs & galaxies & galaxies \\ \hline
E/S0        &   0\% &  27\% & 23\% & 27\%\\
Spiral    &  36\% & 45\%  &  43\% & 70\%\\
LCG  &  25\% & 17\% &  19\% & $<$ 2\%\\
Irregular  &  22\% &  7\%  &  9\% & 3\%\\
Major merger &  17\% &  4\% &  6\% & $<$ 2\%\\ \hline
\end{tabular}
\end{table}

\section{Morphological changes}
\subsection{Galaxy classification}
Distant galaxy morphologies can no longer be described by the Hubble classification sequence: many of them show irregular, complex or compact morphologies (Brinchman et al. 1998; Lilly et al. 1998; Guzman et al. 1997). Table 2 shows the result of a morphological classification performed for 111 $z> 0.4$ galaxies (see also Table 1) for which we have been able to derive their colour maps (see Zheng et al. 2004a for details on the procedure). Here we did not try to disentangle ellipticals from lenticulars; we have counted spirals as a single class and assessed another class (compact) for all objects barely resolved by the HST. From their visual examinations, Brinchman et al (1998) generally classified our compact galaxies as peculiar (irregular \& tadpole), while van den Bergh (2001) classified most of them as spirals. Such a disagreement between the two studies of the same objects leads us to suspect that the classification of compact galaxies is not secure at the WFPC2 spatial resolution. Our very drastic classification scheme based on deep images with two colours has the potential advantage of considerably limiting the uncertainties related to cosmological dimming, spatial resolution and morphological k-correction (see Abraham and van den Bergh, 2001). Apart from compact galaxies, our final classification is in excellent agreement with former ones (Brinchman et al. 1998; van den Bergh et al. 2001).

Following BE2000, let us assume that the CFRS sample is somewhat representative of intermediate mass galaxies between $z=0.4$ and $z=1$, with the morphological type distribution given in Table 2. Morphological determined luminosity functions (MDLF) in the local Universe are available from Nakamura et al. (2004), and one can approximate their mass function using the stellar mass/light ratio in the Sloan r$-$band from Glazebrook et al (2003).  It provides for local intermediate mass galaxies, a number fraction of E/S0 and of early type spirals of 27\% and 53\%, respectively, the rest being mostly late type spirals. It is encouraging to see that, for early type galaxies, the Nakamura et al (2004) MDLF does not differ substantially from that from Marzke et al (1998). It is important to notice that these fractions are barely affected (by a few percent) if we extend the lower mass limit towards a higher value. The most remarkable trend is the small fraction, in the Nakamura et al sample, of peculiar galaxies such as irregulars, LCGs or mergers. An important goal of this paper is to approximately relate the distribution of morphological types of Table 2 to those found in the local Universe. The most significant changes are, by decreasing order of significance, the complete vanishing of LCGs in the local Universe (factor = 10 according to Jangren et al. 2004; see also Garland et al. 2003) and the number decrease of mergers and Irr (factor $\sim$ 4). The fraction of E/S0 appears reasonably constant with the epoch in agreement with Smith et al (2004) and the very modest increase of the E/S0 fraction (also noticed by BE2000) is probably mainly due to the CFRS lack of completeness in measuring redshifts of red galaxies (see section 4.2).

\subsection{Strong star formation in central cores of $z> 0.4$ galaxies}

An important support for the idea of a recent evolution is also provided by the dramatic change in the galaxy central colours. Present-day galaxies in the Hubble sequence exhibit a red bulge superimposed on a blue disk. This is not longer the case for $z > 0.4$ galaxies. Figure 5 shows the distribution of their colours in the central 1 kpc region. A full methodological description can be found in Zheng et al (2004a), and includes the use of  V and I images superimposed with an accuracy of 0.015 $arcsec$, a prerequisite for studying the colour of galaxies with complex morphologies. Almost all distant galaxies (including all LIRGs, Zheng et al. 2004a) have much bluer central colours than local bulges, with the noticeable exception of distant elliptical galaxies, which show similar colours to local ellipticals. We interpret this as evidence for an active star formation in the 1 kpc central region of most distant spirals, which can be either associated with star formation in their bulges, or with an absence of bulge at high redshift, or a combination of both effects. This supports a relatively recent formation of bulges (see Zheng et al., 2004b) in many present-day spirals which are mostly early type.

\begin{figure*}
\begin{center}
\includegraphics[width=0.8\textwidth]{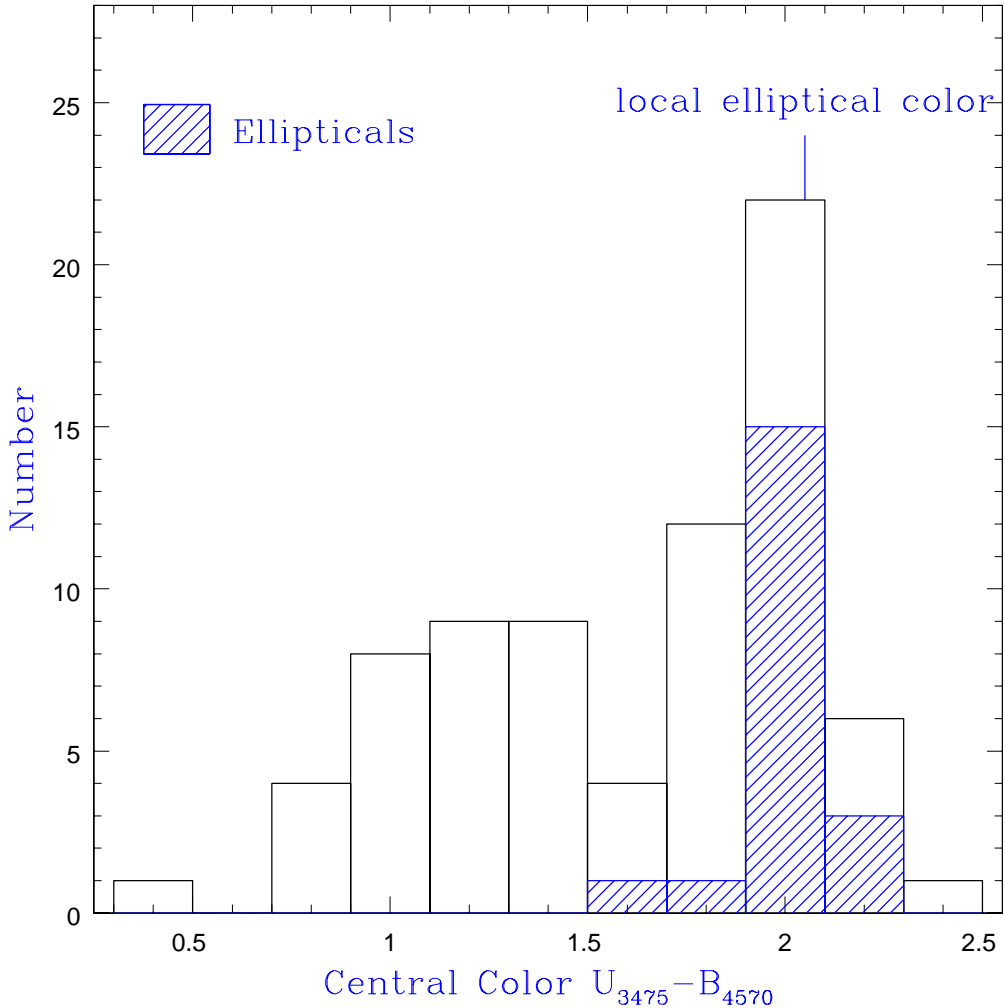}
\end{center}
\caption[]{ Histogram of the 1 kpc inner central color, following Zheng et al. (2004a), for $z > 0.4$ galaxies including ellipticals (shaded area). Central colours have been calculated at $z=0.75$, to minimize k-corrections errors (Hammer et al. 2001). The typical colour of a local bulge at $z=0.75$ (assuming a passive evolution)  is indicated. Morphologically selected ellipticals show no evidence for evolution while the core of most other galaxies are much bluer than expectations from a bulge passive evolution. }
\label{eps5}
\end{figure*}

\section{Towards a scenario of a recent bulge and disk formation in intermediate mass spirals}
\subsection{The need to up-date the current scenario of galaxy formation at $z< 1$}
 Recent observational progress has lead to identification of more than a few rather robust features of the evolution of intermediate mass galaxies. Up to now, no unanimity has been reached in interpreting them within a single picture.
A "galactic downsizing" picture (Cowie et al, 1996; BE2000) requires significant improvements, since apart from a recent and active formation of dwarves, it assumes that most galaxies have completed the bulk of their stellar mass at $z=1$. Its major weaknesses are that:

\begin{itemize}
\item the stellar mass formed since $z=1$ (from 30 to 50\% of the total) is unlikely produced in dwarf galaxies, because these
  objects contribute only a little to the local stellar mass or metals
  (Fukugita et al. 1995; H2004).
  \item it neglects the role of dust-enshrouded star formation, the
  rate density of which has been found to be higher (Flores et al.
  1999) than that derived from rest-frame UV light, and which is also
  required for matching the stellar mass evolution after integrating
  the CSFD (see Dickinson et al. 2003).
\item it is not consistent with the reported deficiency of metal in
  the gas phase of intermediate mass galaxies at $z> 0.4$ (Figure 4).
\item it predicts that intermediate mass spirals and ellipticals were
  mostly evolving passively since $z=1$, while from pair counts, a
  significant fraction of L* galaxies should experience major merger
  events between $z=1$ and $z=0$ (section 1.2).
\item it does not easily account for the large number of peculiar
  galaxies (LCGs, irregulars \& mergers) which represent more than a third of the intermediate
  mass galaxies at $z> 0.4$, and display the most remarkable changes in galaxy
  morphology during the last 8 Gyr (Table 2).
  \item it is qualitatively at odds with the predominance of blue cores
  in the distant galaxies which are not ellipticals.
\end{itemize}

Of course many of the points above have been widely discussed in the literature, and some could have been partly solved in the framework of the "galactic downsizing" picture. Here we argue that the simultaneous changes in galaxy morphologies, central colours, stellar mass production and metal abundance properties are not fortuitous, and we propose to update the current scenario of galaxy formation during the last 8 Gyrs. To do this, we need to puzzle out the impact of the star formation history on the galaxy morphologies and to better understand how distant galaxies are connected to the present-day Hubble sequence.\newline

\begin{figure*}
\begin{center}
\includegraphics[width=1.0\textwidth]{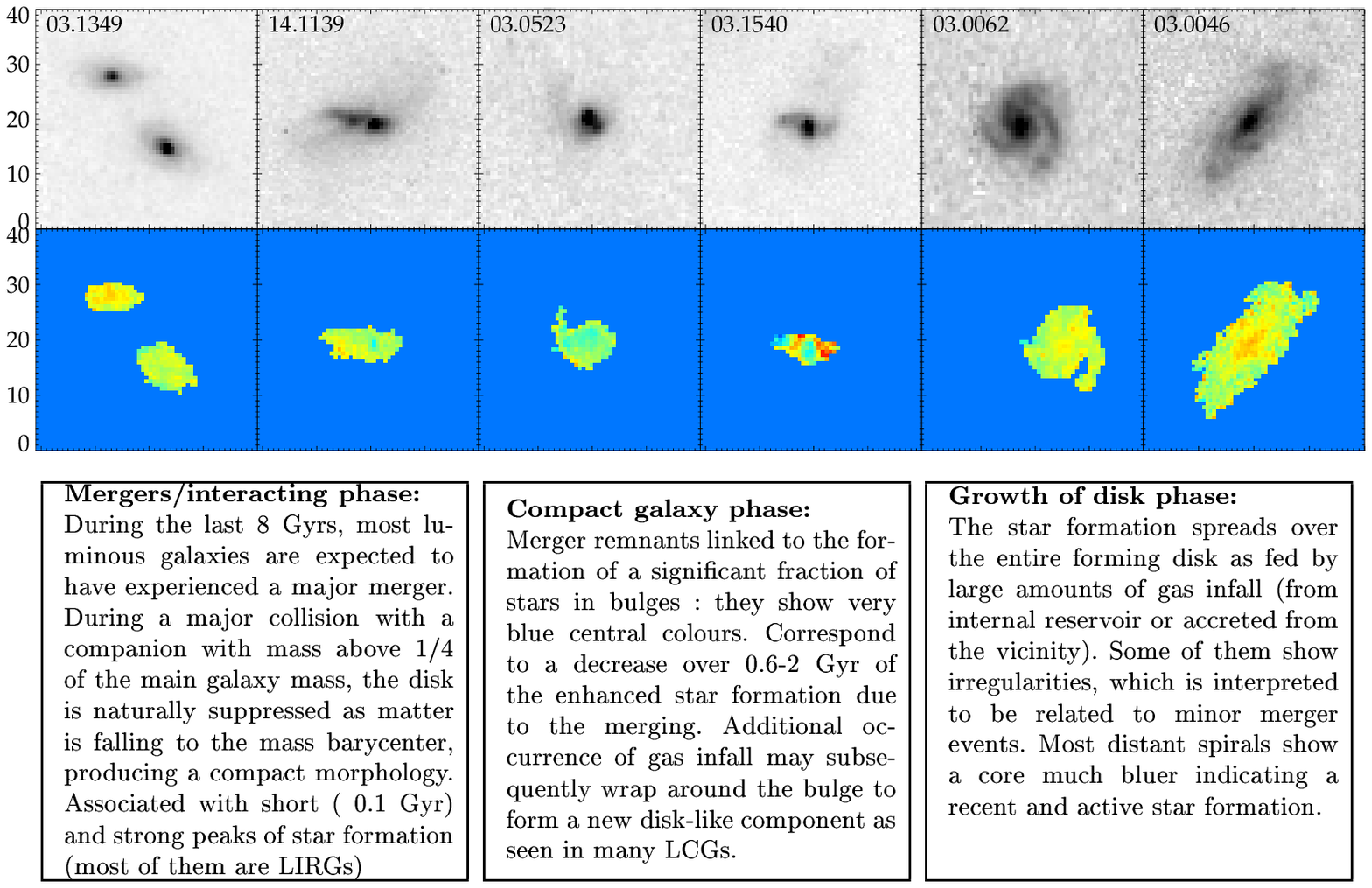}
\end{center}
\caption[]{ A sequence of $z> 0.4$ galaxies illustrating our proposed scenario for a recent formation of spirals in 3 major phases (major merger, compact galaxies, growing disks)}
\label{eps6}
\end{figure*}

\subsection{Where and how have most stars formed since z=1 ?}

 Almost all the recent star formation has occurred in spirals, LCGs, irregulars and major mergers, while the contribution of E/S0 is negligible, both from UV (Lilly et al., 1995; Brinchman et al., 1999; Menanteau, Abraham \& Ellis, 2001; Wolf et al., 2004) and from IR lights (Zheng et al., 2004a; see also Table 2). In which present-day galaxy type has star formation occurred ? Some of the major mergers (those having consumed all their gas) and some of the LCGs possibly gave rise to a recent formation of a few intermediate mass ellipticals. However if the fate of all LCGs (or all LIRGs) were to become ellipticals, the density of ellipticals today would be much higher than observed (see Table 2). Indeed many LIRGs show disk morphologies, many LCGs show faint surrounding disks, and so we assume that most of the recent star formation has occurred in progenitors of the numerous population of present-day spirals. 

The episodic star formation seen from IR data (see section 4) is sufficient to explain most of the stellar mass formed since z=1, and is highly consistent with predictions from hierarchical galaxy formation models. In our view, LIRGs display the most active changes in galaxy morphology, while other galaxies are rather witnessing a more secular galaxy evolution. Zheng et al. (2004a) show that LIRGs present a well-defined sequence in the compactness versus central colour plane, from very blue compact cores (the LCGs) to grand design spirals with central colour consistent with present-day bulges. This observation is taken as the starting point to develop our galaxy formation scenario. Figure 6 illustrates an evolutionary sequence which assumes that several present-day spirals have been shaped during the recent formation of stars and metals in LIRG episodes. Indeed this sequence ends by an inside out formation of the disk, occurring from the outskirts of an already existing bulge. What we are proposing here is that, provided that new gas is available to the galaxy (in an internal reservoir or accreted from the vicinity), it re-emerges as a spiral.

In their detailed simulations of the formation of a single spiral galaxy, Abadi et al. (2003) have assumed a relatively high redshift ($z=1.5$) for the last major merger event, which produces a short and intense starburst episode as well as a complete re-shaping of the galaxy bulge and disk. Our scenario is consistent with a more recent occurrence of the last major merger in most intermediate mass spirals. Galaxy mergers are very efficient in activating star formation as shown by hydrodynamical models, and the (episodic) star formation history of our sequence can follow what it is displayed in Figure 1 of Tissera et al (2002). Major mergers also induce a shock-heated gas wind and the formation of a remnant gas halo, which could have a cooling time of more than one Gyr and could provide an enriched medium with which a gas disk could subsequently be formed (Cox et al, 2004). In the following we will test to what extent a scenario of "spiral rebuilding" is able to reproduce the evolutionary trends of intermediate mass galaxies.

\begin{table*}
\caption {Characteristic times that spiral progenitors spend at each phase from z=1 to z=0.4 ($t_{elapsed}$ = 3.3 Gyrs) }
\begin{tabular}{l|ccccl} \hline
Phase  & observed & characteristic time & characteristic time & theoretical &\\ 
 & fraction    & (= fraction $\times$ 3.3 Gyr)  & " "  & expectations   & References \\ 
 &  & (all spiral involved) &  (75\% of spiral involved) &  & \\
\hline  
Major merger (*)        &   8\%  & 0.26 Gyr & 0.35 Gyr  & 0.2-0.5 Gyr & (1)\\ 
 &  &  &  &  & \\ 
 Compact (major  & 24.5\% & 0.8 Gyr & 1.1 Gyr & 0.6-2 Gyrs & (1), (2), (3)\\
merger remnant) &  &  &  &  & \\
Irregulars (minor  & 11.5\% & 0.38 Gyr & 0.38 Gyrs & 0.2-0.5 Gyrs & (1)\\
 mergers) &  &  &  &  & \\
Spirals & 56\% & 1.8 Gyr & -- & -- & --\\
\\ \hline 
\end{tabular}
\begin{list}{}{}
\item[ $-$] (*) corresponds to close pairs for which the two nuclei are separated by less than one galactic radius 
\item[ $-$] (1): Tissera et al (2003); (2) Baugh et al (1996); (3) Cox et al (2004) 
\end{list}
\end{table*}

\subsection{Ingredients for an updated scenario of "spiral rebuilding": consistency checks and predictions}

 The sequence displayed in Figure 6 means that major mergers/compact/spirals \& irregulars correspond to the various phases undergone by spiral progenitors from z=1 to z=0.4. Let us also assume that the characteristic times corresponding to each phase can be approximated by the number fraction displayed in Table 2 (column 3). Table 3 shows these fractions rescaled after excluding E/S0, as well as the expected characteristic times if all (or alternatively 75\% of) present-day spirals were formed at z$<$1  following our evolutionary sequence. The derived characteristic times match rather well with theoretical expectations based on simulations of major mergers. Our evolutionary sequence is then consistent with the observed fractions of the different galaxy types. Encouraged by this, we also consider the constraint from pair statistics (Le F\`evre et al., 2000, revisited by Bundy et al., 2004, see section 1.2). Assuming that 75$\pm$25\% of galaxies experienced a major merger since z=1 provides a particularily good match of the timescales in Table 3.

Table 4 presents a score card showing different scenarios, including "galactic downsizing" and our scenario ("spiral rebuilding"), and compares their relative merits in reporting eight features of galaxy evolution since z=1. By construction our scenario is consistent with the CSFD evolution derived from UV and IR lights, with the IR background, and with the merging rate evolution. It also reports for 30\% of the current stellar mass formed over the last 8 Gyrs, which is essentially formed in gas rich spirals, those having approximately doubled their masses since $z=1$. This is particularily consistent with the evolution of the L-Z relationship (Figure 4). The blue cores of distant spirals are also easily reproduced if most spirals were recently rebuilt through major mergers. The proposed evolutionary sequence is efficient in producing large bulges, and predicts a number density of late type spirals which sharply decreases from $z=1$ to $z=0$ in agreement with the observations (see Lilly et al. 1998). The relatively small fraction of present-day late type spirals ($\sim$ 20\%) could then correspond to spirals which have not experienced a recent major merger. Although more qualitatively, the relative fraction of LCGs and major mergers selected from optical and IR wavelengths can be also interpreted: there are more optically selected LCGs than mergers (1st column of Table 2) because the merger remnant time-scale is larger than that associated with the collision (see Table 3 and Tissera et al, 2002); however LIRGs include an almost equal number of LCGs as mergers (2nd column of Table 2) because of the lower star formation expected in the remnants. Finally a "spiral rebuilding" scenario is also consistent with an almost non-evolving spiral number density. Some major mergers -but not all- could involve 2 intermediate mass spirals, in which case a decrease of the spiral number density is expected. However this could be easily compensated for by merging of some smaller units whose end-products could enter the intermediate mass range. 

\begin{table*}
\caption { Test of three scenarios  }
\begin{tabular}{l|cccl} \hline
Scenario/Evolution of   & "galactic downsizing" & "spiral rebuilding" & "collisional starbursts"  &\\  & Intermediate mass                   & Minor/major mergers  & Mostly minor merging                        & References \\   
                   &    galaxies already formed at z=1 & 75\% of Sp reprocessed         &             &  \\ 
                    &  (BE2000, Cowie et al,1996)           &   (our scenario/this paper)                            & (Somerville et al, 2001)                 & \\ \hline  
Mass growth         &          {\Large $-$ }               &{\Large  $+$ }                  &{\Large  $+$ } & H2004\\ 
                    &                                      &                                &               & Dickinson et (2004)\\    
L-Z (O/H)           &    {\Large $-$ }                     &{\Large  $+$ }                  &{\Large  $+$ } & Liang et al (2004b)\\ 
                    &                                      &                                &               & this paper\\  
Pair statistics     &    {\Large $-$ }                    &{\Large $+$ }                  &{\Large  $-$ } & Le F\`evre (2000)\\ 
                    &                                      &                                &               &  Bundy et al (2004) \\  
IR light evolution  &    {\Large $-$ }                       &{\Large  $+$ }                  &{\Large  $-$ } & Flores et al (1999)\\
                    &                                      &                                &               &  Elbaz et al(1999) \\  
\# density of E/S0  &      {\Large  $+$ }                  &{\Large  $+$ }                  &{\Large  $+$ } & Schade et al (1999) \\ \\  
\# density of spirals   &      {\Large $+$ }                  &{\Large $+$ }                   &{\Large  $+$ } & Lilly et al (1998) \\  \\   
\# density of peculiar &    {\Large  $?$ }                   &{\Large $+$ }                  &{\Large  $?$ } & Brinchman (1998) \\
(incl. LCGs) & & & & this paper\\
spiral core colours   &      {\Large $-$ }                  &{\Large $+$ }                   &{\Large  $-$ } & Ellis et al (2001)\\  & & & &  Zheng et al (2004a,b)\\
\end{tabular}
\begin{list}{}{}
\item[ $Note:$] a passing grade (+) means that the observables are consistent with the scenario, a negative grade (-) that there seems to be a significant inconsistency and a question mark indicates some doubt about the scenario consistency. 
\end{list}
\end{table*}

 An important uncertainty is related to the actual merger rate since z=1, which requires accurate velocity measurements of both components. At z$\sim$ 0.3, Patton et al (2002) found that one fourth of interacting galaxies are unlikely to be merging systems since their velocity difference exceeds 500 $km s^{-1}$. Changes in the merger rate fraction would affect the relative fraction of (LCGs and LIRGs) events related to either major mergers or tidally triggered starbursts by galaxy interactions. This would not affect the crux of our scenario (Figure 6 still stands), i.e. an episodic star formation history for intermediate mass galaxies, which leads to major changes of their morphologies as expected in hierarchical scenarios. If the major merger event per galaxy was significantly lower than 75\%, it would imply that minor mergers and collisionless galaxy interactions are efficient enough to account for the strong and episodic star formation seen from IR observations. This "soft" alternative of our scenario requires a very (extremely?) large gas content in distant galaxies.   

Table 4 also tests a third scenario (called "collisional starburst") assuming a marginal impact of major mergers. Such a scenario has been developped by Somerville, Primack \& Faber (2001), and does not account for collisionless galaxy interactions. It succeeds in interpreting many galaxy evolution features up to z=3, but failed in reporting the star formation densities derived from ISO and SCUBA experiments. It would also be difficult to reconcile it with the numerous populations of distant compact galaxies and with the blue cores of most distant spirals.

\begin{figure*}
\begin{center}
\includegraphics[width=0.8\textwidth]{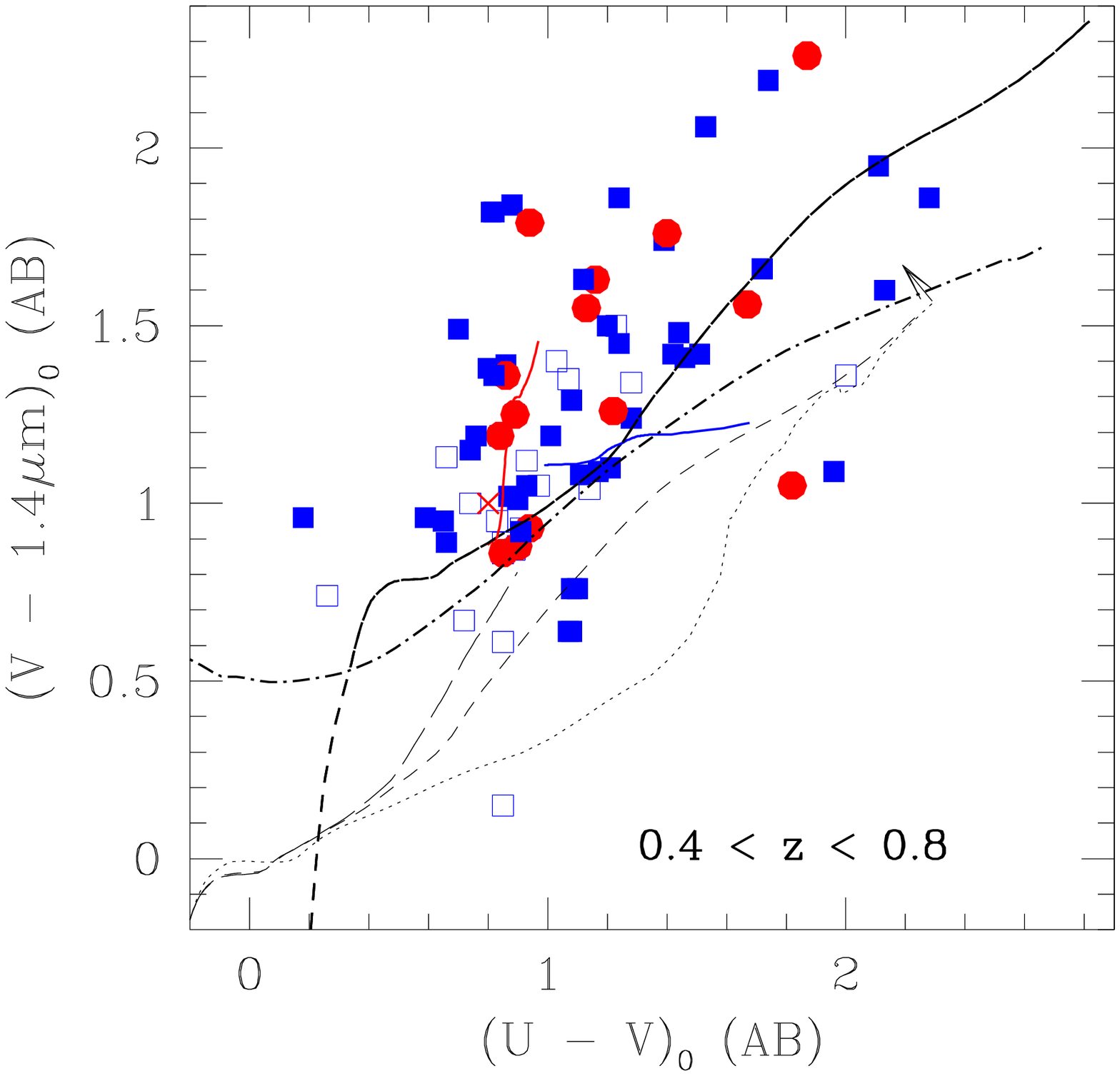}
\end{center}
\caption[]{ Rest-frame color $(V - 1.416 \mu$m$)_{AB}$ versus 
$(U - V)_{AB}$ for CFRS luminous galaxies ($M_{B} \le$ -20, same symbols as in  Figure 4). It is adapted from Hammer et al. (2001, see their Figure 5), who have shown that the rest-frame colours used here match particularily well the observed colours in the redshift range z=0.4-0.8, leading to very small errors related to k-corrections. Here we represent CFRS galaxies for which the photometric error bars on colours are below 0.4 magnitude. 
 Lines show Bruzual and
Charlot (1999) models with solar metallicity and an exponential
declining SFR with an e-folding time-scale $\tau$.  The dotted line
show the model with $\tau$ = 0.1 Gyr, the short-dashed line the model
with $\tau$ = 1 Gyr and the long-dashed line the model with a constant
SFR. The bold short-dashed line represents the $\tau$=1 model with an
extinction $A_{V}$ = 1 applied. The bold dot-short dashed line
represents the $\tau$ = 1 Gyr model, assuming a metallicity
$Z/Z_{\odot}$ = 3.2.  A composite model, where 80\% of the light is
contributed by an old stellar population formed 5 Gyr ago in an
instantaneous burst, and the remaining 20\% of the light by a younger
stellar population (from t=0 to t=1Gyr) with $\tau$ = 1 Gyr, is shown as a solid (blue, almost horizontal) line.
Red (almost vertical) line show the superposition of a normal starburst color (cross) with an extincted starburst (see text).  A small arrow on
the upper-right side of the Bruzual and Charlot (1999) models
indicates the uncertainties in the modeling of the old stellar
population, as discussed by Charlot et al. (1996).}
\label{eps7}
\end{figure*}

\section{Discussion}
To firmly determine the above scenario will be very demanding of observing time, requiring deep multi-wavelength observations (photometry \& spectroscopy) for a much larger sample than the one presented here.  Our results provide a good support for the hierarchical galaxy formation framework since most of the star formation is occurring episodically; at the level of our modest statistics, we predict that 42, 22 and 36\% of the IR (episodic) star formation density is related to major mergers, minor mergers and gas infall, respectively. On the other hand, detailed simulations are also needed, especially to derive more accurate time-scales, and precise the mechanisms controlling the disk growing (Birboim and Dekel, 2003; Dekel and Birnboim, 2004, in preparation). Given the fact that our scenario links distant and local galaxy properties in a simple way, we investigate here some of its potential failures.\newline

A major problem in reporting evolutionary trends of galaxy evolution from all inclusive surveys is related to Malmquist biases, which severely limit their statistical significance. Indeed all inclusive surveys particularly are not adapted to follow galaxy evolution in the frame of hierarchical models, because the bright (generally massive) objects seen in the highest redshift bin are unlikely the progenitors of fainter (less massive) galaxies seen at lower redshifts. Here we simply consider all intermediate mass galaxies selected in a high-z volume ($0.4 < z < 1$) and then interpret their changes after relating them to local galaxies in the same mass range. Clearly our scenario prediction (most galaxies have recently doubled their masses) introduces a bias when comparing to local galaxies, but we find it produces a marginal effect, since the relative fractions of present-day E/S0 and spirals are found barely dependent on changes in the low mass cut-off.\newline 
 
One can also cast doubt on the fact that the CFRS I-selected sample is representative of all intermediate mass galaxies in a given redshift range. Our conclusions indeed account for a significant lack of red quiescent galaxies in the CFRS. Also, could we have missed a significant population of emission line galaxies ? Figure 7 shows the rest-frame colour-colour diagram for our sample. Independently of their nature (LIRGs, LCGs \& normal starbursts), most galaxies are found within ($(U-V)_{0}$ = 1 $\pm$ 0.4 and $(V-1.4\mu m)_{0}$ = 1.3 $\pm$ 0.5. There is only a hint that normal starbursts show bluer colours than LCGs and LIRGs, which is not unexpected if they are less affected by dust extinction. It means that a selection of galaxies using K band (rest-frame 1.4$\mu m$) or I band (rest-frame V) should find essentially the same mixture of emission line galaxies. Using the CFRS I-band limit we might have missed only a few dusty enshrouded starbursts which are similar to the LIRGs and LCGs at the top of Figure 7. \newline 

Have we overestimated our stellar masses and misinterpreted a possible population of low mass galaxies in our sample ?  Stellar masses are estimated using various schemes for galaxy star formation histories, generally through a combination of a (quiescent) underlying population with a burst (see Figure 7). Indeed our estimates are in the same range of those of BE2000 (same sample), agree within 30\% with Franceschini et al. (2002) for LIRGs, and are in good agreement with those of Guzman et al. (2003) for LCGs (which correspond to their brightest sources in the HDF). However, the main uncertainty in the inferred stellar masses arises from the age of the stellar population, which leads us to suspect that masses of the most active star-forming systems (e.g LIRGs) possibly have been overestimated relatively to those of more quiescent galaxies. This effect has been thoroughly investigated by Charlot (1998) who finds values of $M/L_{K}$ as low as 0.2 for 0.1 Gyr old pure starbursts. LIRG spectral energy distributions (from UV to mid-IR and radio when available) are consistent with the superposition of a young and highly reddened burst to a stellar population found in local spirals (Flores et al. 1999). Assuming a 0.1 Gyr burst corresponding to the formation of 20\% of the mass would lead to decrease the $M/L_{K}$ ratio for LIRGs from $\sim$1 to 0.5, accounting also for the modest effect of extinction at that wavelength. The effect does not severely affect the fraction of LIRGs in intermediate mass galaxies (it drops from 15\% to 12\%), and does not affect the overall mass production related to LIRGs. However the median doubling time-scale is then reduced to 0.4 Gyr, leading to the formation of $\sim$ 30\% of the mass during a single infra-red event. A more rapid star formation in some galaxies with masses just below 3 $10^{10} M_{\odot}$ would have also the advantage of more naturally relating distant galaxies to present-day intermediate mass galaxies.  The detailed modelling of the stellar features of the VLT-FORS2 moderate resolution spectra of distant LIRGs will provide strong constraints on their past star formation histories, the intensity of the burst, its duration and the stellar mass fraction born during the burst (Marcillac et al. 2004). 


 We found that our results (morphological \& spectral properties, LIRG fraction) are not significantly affected by adopting different schemes in assessing the stellar mass or by using different broad band selection criteria for our high-z sample. All $M_{B}< -$20 distant galaxies in the CFRS are found to be intermediate mass galaxies whether they are quiescent, "normal" starbursts, LCGs or LIRGs. 
In fact, we find that galaxy continuum properties (absorption line spectrum \& broadband colours) are mostly unaffected by the burst strength, which can be simply understood if the burst brightening is somewhat balanced by the effect of dust extinction. In other words, {\it optical properties of most LIRGs mimic those of normal starbursts}. We conclude that selecting z$<$ 1 galaxies from their apparent I to K band flux does not lead to a significant bias against a given class of galaxies, and that, to first order, what we see from these selections is what is there. This is at odds with the idea that present-day dwarf galaxies were playing a significant role in high z samples (Cowie et al. 1991; Koo et al. 1992) simply by a huge burst-brightening in optical. Besides the fact that this idea does not match with distant galaxy spectra and colours, another argument against the dwarf hypothesis is provided by the metal abundances of distant galaxies: although lower than typical values for intermediate mass spirals, these are much larger than expectations from progenitors of present-day dwarves. As an aside, one early motivation for our study of distant LCGs was to detect the [OIII]$4363$ line which is $\sim$1/3 the intensity of the $H\gamma$ line in local poor metal galaxies. No such detection (except for one source) has been made using deep exposure spectroscopy with the VLT, and the LCG spectra have conversely revealed a continuum with strong absorption lines, as expected from old stellar component. \newline

An important assumption of our scenario is the requested availability of supplementary gas to progressively form the stellar disk content after the compact phase. Much higher gas infall was expected in the past as shown by the large redshift increase of the number densities of starbursts and LIRGs. A high gas reservoir in progenitors of spirals is not unexpected at these epochs, because 10 Gyr ago ($z=2$), most of the baryonic mass was in the form of ionised gas, whose mass content has by now dramatically decreased, by a factor $\sim$ 50, at $z=0$ (Fukugita, Hogan and Peebles, 1998), even when accounting for neutral and molecular gas. We argue that this change of state of the baryonic matter did not take place only between $z=2$ and $z=$1 (elapsed time 2.6 Gyr), but is likely the result of a longer process over the period between $z=2$ and $z=0.4$ (elapsed time 6 Gyr). Cosmic chemical evolution models (Pei, Fall and Hauser, 1999) successfully reproduce the CSFD evolution, and assume a decrease of the ionised gas density by a factor 5 since $z=1$. Indeed in the distant large spirals that are LIRGs, the star formation process is spread  all over the disk, as shown by its blue colour (Zheng et al. 2004a), as expected if large amounts of gas are being accreted. \newline




\section{Conclusion}

 We have re-examined here the properties of distant galaxies mostly on the basis of a panchromatic follow-up study of the CFRS. We find that:
\begin{itemize}
\item SFRs based on the [OII]$\lambda$3727 emission line lead to severely underestimated values, which can lead to misleading results on the star formation history of distant galaxies
\item 15\% of intermediate mass galaxies at 0.4$<$ z $<$ 1 are indeed LIRGs and the associated star formation density is sufficient in itself to form 38\% of the total stellar mass in present-day intermediate mass galaxies
\item all emission line galaxies (starbursts and LIRGs) at z$\sim$ 0.7 have gas phases which are metal deficient by a factor 2 when compared to those of local spirals
\item most distant field galaxies (but E/S0) show significant star formation in their 1kpc radius central cores
\item  optical/spectral properties of LIRGs are so similar to those of other galaxies that only IR measurements are able to describe how the star formation density is distributed between the different galaxy types
\item the only way to interpret the high occurrence of LIRGs is to associate them to episodic star formation histories for most galaxies, in agreement with the hierarchical galaxy formation scenario   
\end{itemize}

 These episodic and violent star formation events can be simply associated to major changes in galaxy morphologies. We have then tested three galaxy formation scenarios against eight rather robust features of galaxy evolution. The best scenario accounts for a recent formation of bulges in spirals through merging or tidally triggered central bursts, followed by an inside out formation of the stars in disks. In this scenario, at least half of the bulge stellar content was made earlier in their progenitors, before the last major phase of accretion. It then provides a simple origin for the reported recent stellar mass production (30 to 50\% of the stellar mass formed since z=1) and for a recent peak in the cosmic star formation history (H2004). It is consistent with the simultaneous decreases, during the last 8 Gyrs, of the UV and IR luminosity density, of the merging rate, and of the number densities of LIRGs and of LCGs. It also naturally accounts for all the major changes of the galaxy morphologies in the last 8 Gyrs, as well as for an approximately unchanged density of large disks at $z\sim 1$. Because of the striking similarities between properties of galaxies at different luminosities (average $O/H$ abundances, fraction of merging events and of compact galaxies), possibly most present-day spirals, independently of their mass range, have recently experienced their last major phase of the building of both their bulges and their disks. More than a third of the present-day stellar mass formed at $z < 1$ is associated with more than half of the stellar mass formed in spirals during violent events which have shaped them as the Hubble sequence describes them today.\newline

 All the important numbers used here to establish our scenario (e.g. fraction of LIRGs and of various morphological types, SFR, stellar masses etc...) have been systematically cross-checked and found in good agreement with the litterature. We also find that biases related to the I band selection criterion used in the CFRS provide minor effects. This is likely due to the fact that dust-enshrouded starbursts (such as LIRGs) show optical properties which mimic those of normal starbursts, in such a way that optical observations are generally unable to distinguish them. This is well reported in our scenario: the addition of a highly reddened burst to a normal high-z galaxy has no severe impact on its optical continuum. The way IR light impacts our optical view of the galaxy evolution is then rather subtle and it explains well the often associated wording "hidden star formation". Without observing in the IR and accounting only from the UV light, all distant galaxies apparently show modest star formation rate and stellar mass increase. 
 \newline     

   A significant number of recent major mergers is qualitatively consistent with the dramatic increase of type 1 and 2 AGN out to z=1 reported by Miyaji, Hasinger \& Schmidt (2000).  Major mergers are particularily efficient in forming large bulges which can account for the numerous population of present-day early-type spirals (see also Zheng et al, 2004b). The link between distant and nearby galaxies could be further investigated: what is the relative impact of both star formation and mass aggregation at z$<$ 1 for present-day spirals and ellipticals ? Could we match our scenario with the Milky Way and M31 past histories ? While it is unlikely that a major merger has recently occurred in the Milky Way, the M31 system appears to be much more complex and the significant intermediate age population in its halo supports the idea of a recent galaxy merger (Brown et al, 2003). It is also necessary to predict more accurately the relative fraction of star formation occurring during major mergers, minor mergers, interactions and gas infall, respectively. This requires unambiguous estimates of the actual merging rate and of the fraction of LCGs which are end products of merging events. We believe that estimates of velocity fields of distant galaxies with the VLT (with FLAMES/GIRAFFE) will address these issues, as well as the evolution of the Tully Fisher relationship. Large follow up of the deepest fields studied by HST and Spitzer should be done, using moderate spectral resolution spectroscopy to derive SFR, extinction and chemical gas properties, as well as velocity fields provided by 3D spectroscopy.

\begin{acknowledgements}
This paper has benefitted from enlightening discussions with Matt Lehnert and Fran\c{c}oise Combes and their comments have contributed to improve its content. We are grateful to the anonymous referee for severe (although justified) comments and to Leslie Sage; their remarks were useful in improving our ability to communicate our results. 
\end{acknowledgements}

\end{document}